\documentclass[journal=jacsat]{achemso}
\setkeys{acs}{articletitle = true}

\usepackage[dvipsnames]{xcolor}
\usepackage{color}
\usepackage[version=3]{mhchem} 

\usepackage{hyperref}

\usepackage{times}
\usepackage{amsmath}
\usepackage{amssymb}
\usepackage{bm}
\usepackage{here}
\usepackage{graphicx}
\usepackage{framed}
\definecolor{shadecolor}{rgb}{0.9,0.9,0.9}
\usepackage[utf8x]{inputenc}

\usepackage{lipsum}

\usepackage{babel}
\usepackage{amsmath}
\usepackage{amssymb}
\usepackage{bbold}
\usepackage{wasysym}
\usepackage{graphicx}
\usepackage{multirow}
\usepackage{gensymb}
\usepackage[dvipsnames]{xcolor}
\usepackage{pifont}
\usepackage{microtype}
\usepackage{physics}
\usepackage{xcolor}
\usepackage{soul}
\usepackage[version=3]{mhchem}
\usepackage{chemformula}[2014/04/08] 
\usepackage{booktabs}    
\usepackage{dcolumn}     
\usepackage{bm}          
\usepackage{endnotes}

\newcommand{\vect}[1]{\ensuremath{\bm{#1}}}

\author{Francesco Belli}
\affiliation[State University of New York at Buffalo]
{Department of Chemistry, State University of New York at Buffalo, Buffalo, NY 14260-3000, USA}
\author{Eva Zurek}
\affiliation[State University of New York at Buffalo]
{Department of Chemistry, State University of New York at Buffalo, Buffalo, NY 14260-3000, USA}
\email{ezurek@buffalo.edu}
\author{Ion Errea}
\affiliation[University of the Basque Country (UPV/EHU)]
{Fisika Aplikatua Saila, Gipuzkoako Ingeniaritza Eskola, University of the Basque Country (UPV/EHU), Europa Plaza 1, 20018 Donostia/San Sebasti\'an, Spain}
\alsoaffiliation[Centro de F\'isica de Materiales (CSIC-UPV/EHU)]
{Centro de F\'isica de Materiales (CFM-MPC), CSIC-UPV/EHU, Manuel de Lardizabal Pasealekua 5, 20018 Donostia/San Sebasti\'an, Spain}
\alsoaffiliation[Donostia International Physics Center (DIPC)]
{Donostia International Physics Center (DIPC), Manuel de Lardizabal Pasealekua 4, 20018 Donostia/San Sebasti\'an, Spain}
\email{ion.errea@ehu.eus}

\DeclareMathAlphabet\mathbfcal{OMS}{cmsy}{b}{n}
\newcommand{\tc}{$T_\text{c}$}

\title{A chemical bonding based descriptor for predicting the impact of quantum nuclear and anharmonic effects on hydrogen-based superconductors}

\begin{document}

\clearpage
\newpage
\begin{abstract}
Quantum nuclear effects (QNEs) can significantly alter a material's crystal structure and phonon spectra, impacting properties such as thermal conductivity and superconductivity. However, predicting \emph{a priori} whether these effects will enhance or suppress superconductivity, or destabilize a structure, remains a grand challenge. Herein, we address this unresolved problem by introducing a descriptor, based upon the integrated crystal orbital bonding index (iCOBI), to predict the influence of QNEs on a crystal lattice's dynamic stability, phonon spectra and superconducting properties. We find that structures with atoms in symmetric chemical bonding environments exhibit greater resilience to structural perturbations induced by QNEs, while those with atoms in asymmetric bonding environments are more susceptible to structural alterations, resulting in enhanced superconducting critical temperatures.
\end{abstract}

\newpage

\section{Introduction}

Achieving room-temperature and room-pressure superconductivity would benefit society in unforeseeable ways. Hydrogen-rich systems are the most promising compounds that could achieve this feat, as their strong electron phonon coupling (EPC) and large Debye temperatures allow them, in principle, to possess high superconducting critical temperatures, \tc s, as originally proposed by Ashcroft~\cite{Ashcroft1,Ashcroft2}. Additionally, since the pairing mechanism in hydrides is presumed to be conventional, their superconducting properties can often be reasonably well predicted by theory, opening the door towards rational materials design. Indeed, many hydrides with high \tc s have been synthesized, though at pressures impractical for applications. This includes H$_3$S (\tc\ = 203~K near 150~GPa) \cite{Drozdov:2015a}, LaH$_{10}$ (\tc\ = 250-260~K near 200~GPa~\cite{somayazulu2019evidence,drozdov2019superconductivity}), YH$_4$ (\tc\ = $\sim$88~K at 155~GPa~\cite{shaoHighpressureSynthesisSuperconducting2021}), YH$_6$ (\tc\ = 220~K at 160-180~GPa~\cite{troyan2021anomalous,kongSuperconductivity243Yttriumhydrogen2021}), YH$_{9}$ (\tc\ = 240~K at 200~GPa~\cite{kongSuperconductivity243Yttriumhydrogen2021}), CaH$_6$ (\tc\ = 210-215~K at 160-172~GPa~\cite{Ma:CaH6,Li:CaH6}), (La,Y)H$_{6}$ and (La,Y)H$_{10}$ (\tc\ = 237 and 253~K, respectively, both between 170-196~GPa~\cite{semenokSuperconductivity253Lanthanum2021}), as well as LaBeH$_{8}$ (\tc\ = 110~K at 80~GPa~\cite{songStoichiometricTernarySuperhydride2023a}),  (La,Ce)H$_{9-10}$ (\tc\ = 176~K at 100~GPa~\cite{chenEnhancementSuperconductingProperties2023}) and Y$_{0.5}$Ce$_{0.5}$H$_{9}$ (\tc\ = 97-141~K between 98-200~GPa~\cite{chenSynthesisSuperconductivityYttriumcerium2024}), to name a few. 

As alluded to above, first-principles calculations based on density functional theory (DFT) have proven to be a valuable tool to direct experimental results~\cite{Livas,Pickard,Zurek}, and many of the experimental discoveries have been anticipated by DFT calculations \cite{168,Duan2,Liu2,Peng,liu2017potential}. Furthermore, through DFT-based crystal structure prediction (CSP) methods, the landscape of hydrogen-host binary combinations has been meticulously explored. As a result, a clear understanding of the features enhancing superconductivity in electron-phonon mediated hydrogen-based superconductors has emerged~\cite{Livas, semenok2020distribution, belli2021strong}. Compounds with weakened covalent bonds, elevated hydrogen content, symmetric bonding configurations, and a high density of states (DOS) at the Fermi level ($E_\text{F}$), with a large contribution arising from the hydrogen atoms, are likely to exhibit a high \tc . Such insights are now being applied to direct the \emph{ab initio} exploration of the phase diagrams of ternary and quaternary hydrides~\cite{Kokail,Saha,Sun,DiCataldo,Lucrezi,DiCataldo:2021a,Liang}. The hope is to further expand the list of predicted compounds (including those that are metastable), and scout for low-pressure-synthesizable systems, which would be necessary for practical applications of hydrogen-based superconductors.

The typical workflow~\cite{Zurek:2018m} for the \emph{ab initio} prediction of hydrogen-based superconductors begins with identifying promising structures that correspond to (low energy) local or global minima in the Born-Oppenheimer energy surface (BOES). Next, the dynamic stability of the predicted compounds is determined by calculating the vibrational phonon frequencies in the harmonic approximation, and ensuring that no imaginary eigenvalues are present. This corresponds to a classic treatment of the nuclei, which are assumed to vibrate around their local minima. If, instead, the nuclei are treated as quantum particles, their positions are not clamped, but  fluctuate around an average atomic position, known as a centroid. In this case, the phonon frequencies should be calculated from the second derivative of the total free energy with respect to the centroid positions, which includes the kinetic energy associated to the ionic quantum fluctuations.  In hydrogen rich compounds, the harmonic classical treatment of the nuclei can yield predictions that differ vastly from those obtained with a proper quantum anharmonic treatment of the ions due to the light mass of hydrogen and the dynamical instabilities predicted at the harmonic level because of their potentially very large EPC~\cite{errea2020quantum}. The properties that are affected by quantum nuclear effects (QNEs) and anharmonicity in hydrides include: the optimized geometries, phonon frequencies, and perhaps most significantly for the current study, the superconducting behavior. These effects were shown to be extremely important for the first two high pressure hydrides that were synthesized,  LaH$_{10}$~\cite{errea2020quantum} and H$_3$S~\cite{errea2016quantum}. For example, DFT-based CSP searches found an $Fm\bar{3}m$ symmetry LaH$_{10}$ structure to be the most stable at high pressures, but below 230~GPa other distorted variants were preferred, suggesting a complex BOES with many local minima.~\cite{Peng,liu2018dynamics,liu2017potential,errea2020quantum} However, within the harmonic approximation none of the proposed LaH$_{10}$ phases (including $Fm\bar{3}m$) are dynamically stable below 230~GPa, so their \tc\ cannot be calculated. At the same time, experiments measured a \tc\ of 250~K from 137-218~GPa~\cite{Drozdov:2015a} or 260~K at 188~GPa~\cite{somayazulu2019evidence} for a presumed LaH$_{10}$ stoichiometry compound, resulting in a discrepancy between theory and experiment. This discrepancy could only be solved when QNEs were taken into account~\cite{errea2020quantum}. Since then, it has been shown that QNEs have a profound impact on the stability and superconducting properties of numerous high~\cite{hou2021strong,Hou3,Zurek:2023m} and ambient-pressure~\cite{34,Zurek:2024g} hydrides.

Though a number of computational techniques have been developed for treating QNEs and anharmonicity, the method that is often applied to superconducting hydrides is the stochastic self-consistent harmonic approximation (SSCHA)~\cite{ErreaSSCHA, Raffaello, Lorenzo2, Lorenzo, monacelli2021time}. Recently,  workflows that accelerate SSCHA calculations by leveraging machine-learned interatomic potentials have been developed~\cite{Zurek:2024g,fields2023temperature,lucrezi2023quantum}. Therefore, we can expect the number of SSCHA-based studies on hydrides to grow. However, the general trends of how QNEs and anharmonicity impact the superconducting properties of hydrogen-rich compounds is still not clear. In some cases they largely suppress superconductivity -- a prime example being PdH where the strong anharmonicity of the material leads to an overestimation of the \tc\ by almost a factor of four ($\sim$35~K harmonic vs.\ $\sim$10~K experiment)~\cite{34,ErreaSSCHA}, while in other situations, \tc\ can be considerably increased, such as in LaBH$_8$, where QNEs enhance the \tc\ from about 100~K to 160~K at 100~GPa~\cite{belli2022impact}. 

Herein, by collecting a series of systems where QNEs and anharmonicity were treated through the SSCHA, we show the emergence of two possible scenarios. In the first, when inclusion of quantum nuclear effects does not perturb the geometry of a structure, they lower the \tc. Conversely, when quantum treatment of the lattice changes its geometrical parameters,  a concomitant increase in \tc\ is observed. Furthermore, we develop a chemical bonding based descriptor that can predict which of these two scenarios describe a particular hydride. This descriptor, based upon a vector sum of the integrated crystal orbital bonding index (iCOBI)~\cite{muller2021crystal}, is facile to compute, thereby enabling the automated classification of any arbitrary hydride into one of the two possible types. Our study results in a comprehensive, chemical bonding inspired understanding of the impact of QNEs and anharmonicity on the structures and phonon modes of hydrogen-based compounds, offering the possibility of discerning in advance their impact on the superconducting properties.

\section{Methods}


The structural parameters and Eliashberg spectral functions, both with and without SSCHA-treated QNEs, were obtained from the literature. The dataset includes the following compounds: PdH at ambient pressure~\cite{34}, AlH$_3$ at 135~GPa~\cite{hou2021strong}, LaH$_{10}$ at 250~GPa~\cite{errea7}, a phase of atomic hydrogen ($I4_1/amd$ at 500~GPa~\cite{6}), PtH at 100~GPa~\cite{ErreaSSCHA}, YH$_6$ at 150~GPa~\cite{troyan2021anomalous}, H$_3$S-$Im\bar{3}m$ at 250~GPa and 130~GPa~\cite{errea2016quantum}, ScH$_6$-$P6_3/mmc$ at 140~GPa~\cite{Hou3}, ScH$_6$-$Cmcm$ at 100~GPa, H$_3$S-$R3m$ at 130 GPa~\cite{errea2016quantum}, LaBH$_8$ at 100 GPa~\cite{belli2022impact}, and a phase of molecular hydrogen ($Cmca$-4 at 450~GPa~\cite{190}).  Moreover, additional calculations were performed to obtain data for ScH$_6$-$Cmcm$ at 100~GPa, as described below. All these listed spacegroups and pressures correspond to the classical calculation, excluding the potential impact on the structure and the estimated pressure of NQEs.

In most of these investigations~\cite{ErreaSSCHA,hou2021strong,6,190,troyan2021anomalous,errea7,errea2016quantum,belli2022impact,Hou3}, the geometries were initially optimized within the Born-Oppenheimer approximation at the specified pressures. The SSCHA was subsequently applied to relax the atomic positions including NQEs and anharmonicity, while keeping the unit cell shape and dimensions fixed. This approach allows for a direct comparison of the phonons and the Eliashberg spectral functions at identical volumes; however, the pressures determined via SSCHA are higher from those computed using classical nuclei due to the extra contribution arising from the QNEs. In contrast, the comparative analysis between the classical and quantum structures for ScH$_6$-$Cmcm$ at 100 GPa and ScH$_6$-$P6_3/mmc$ at 140 GPa was conducted by relaxing the structures with the SSCHA also at 100 and 140 GPa, respectively, which implies that the lattices in the comparison are different in these two cases.

The ScH$_6$-$Cmcm$ phase was identified by displacing the atoms according to the eigenvectors of the imaginary $E_{2g}$ symmetry mode at $\Gamma$ computed for ScH$_6$-$P6_3/mmc$ at 100~GPa~\cite{Hou3}, followed by a subsequent structural relaxation. The DFT calculations for ScH$_6$-$Cmcm$ were performed with the plane-wave {\textsc Quantum ESPRESSO} package~\cite{Giannozzi1,Giannozzi2}. We employed the Perdew-Burke-Ernzerhof~\cite{GGA-PBE} parameterization of the exchange-correlation potential (PBE-GGA), along with ultrasoft pseudo-potentials that treated 11 electrons of scandium in the valence, with cutoffs for the wavefunctions and density chosen as 1088~eV and 10,884 eV, respectively. The Brillouin zone integration in the self-consistent calculations were performed with a first-order Methfessel-Paxton smearing with a broadening of 0.27~eV, and a $\mathbf{k}$-mesh with a spacing of 2$\pi\cross$0.011 \AA$^{-1}$. The harmonic phonon calculations were performed on a $\mathbf{q}$-mesh with a spacing of 2$\pi\cross$0.056 \AA$^{-1}$, making use of density functional perturbation theory~\cite{Baroni}.  The electron-phonon interaction was calculated on a $\mathbf{k}$-mesh with a spacing of 2$\pi\cross$0.0083 \AA$^{-1}$, with a Gaussian smearing of 0.11~eV to approximate the Dirac deltas, and with the same phonon grid that was used in the phonon calculations. The \tc\ for all compounds were recalculated using the isotropic Migdal-Eliashberg~\cite{ALLEN19831} formalism with a Coulomb repulsion parameter, $\mu^*$, of 0.1.

To analyze the bonding, DFT calculations for all of the structures in our dataset were performed using the Vienna \emph{ab initio} Simulation Package ({\textsc VASP})~\cite{hafner2008ab}, with the PBE-GGA and an energy cut-off of 800~eV. The number of valence electrons treated explicitly and their electronic configurations are reported in Table S1. The core electrons were treated with the projector augmented wave (PAW) method~\cite{PhysRevB.50.17953}. The reciprocal space was sampled using a $\Gamma$-centered Monkhorst-Pack~\cite{PhysRevB.13.5188} $\mathbf{k}$-mesh with a spacing of 2$\pi\cross$0.016 \AA$^{-1}$ with an electronic smearing of 0.2~eV. The integrated Crystal Orbital Bonding Index (iCOBI)~\cite{muller2021crystal}, which is a quantification of the extent of covalent bond formation, was calculated for atom pairs using the {\textsc LOBSTER}~\cite{nelson2020lobster} code with the standard basis set proposed.

\section{Results and Discussion}

\subsection{Two classes of structures}

We begin by introducing the compounds that will be considered in our foray into the impact of QNEs on the properties of superconducting hydrides. Their structures and computed superconducting parameters have been taken from the cited literature, with the exception of the calculations outlined in the \emph{Methods} section. As we will soon see, each one has its own story of how it is impacted by QNEs. As might be surmised by the reader, the way in which QNEs manifest themselves for the ``Type I'' systems described below differs markedly from how it influences compounds that belong to the ``Type II'' class. 

\begin{figure*}[t!]
\includegraphics[width=\textwidth]{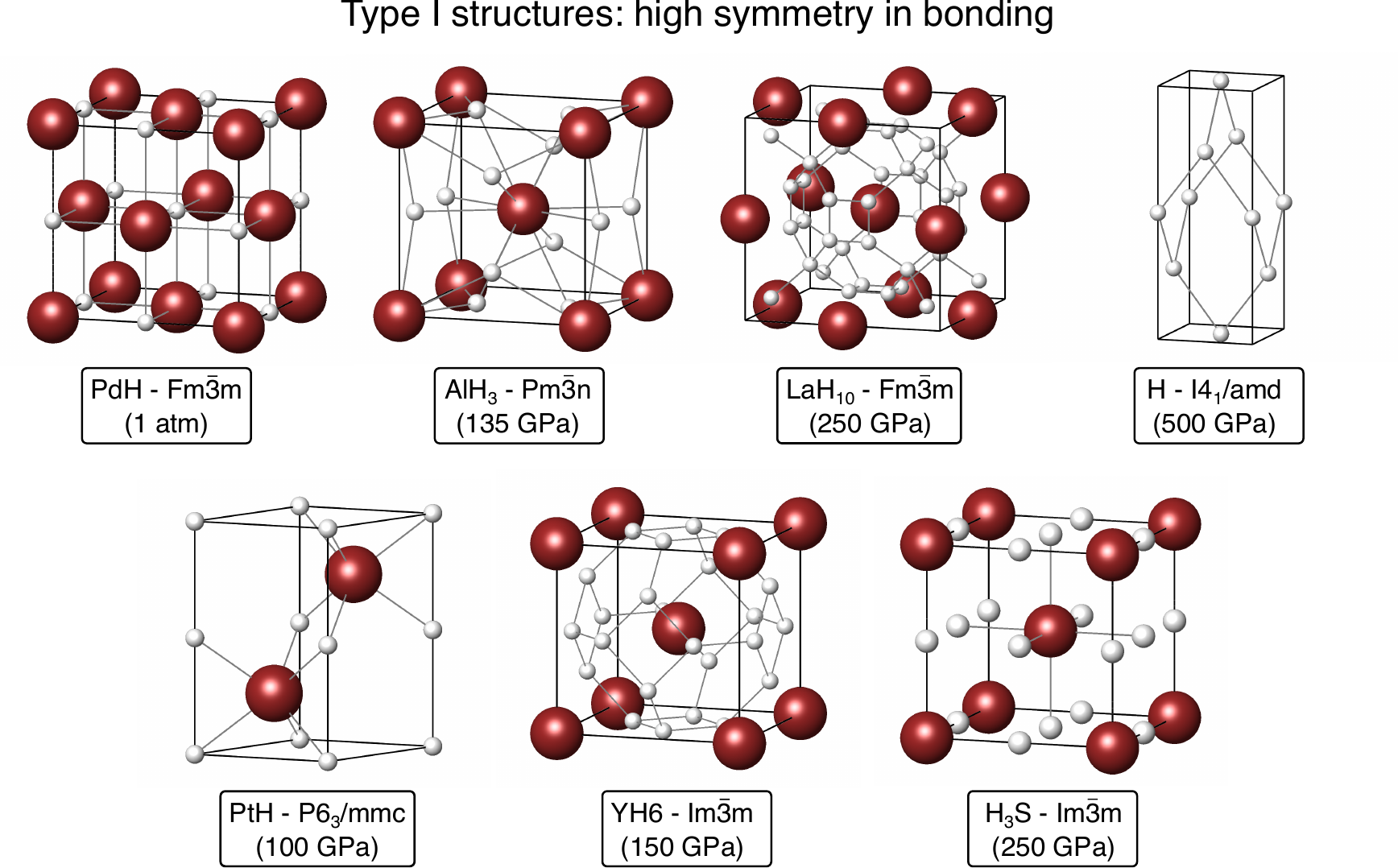}
\caption{\label{Fig:Type1} Crystal structures of the Type I class of compounds. Locally, these phases exhibit a high degree of symmetry in their bonding environments. Hydrogen atoms are white, and all other atom types are colored reddish-brown. }
\end{figure*}

Figure \ref{Fig:Type1} illustrates the Type I crystal lattices and provides the pressure at which they have been computationally investigated. It turns out that the high symmetry of these lattices, and absence of free parameters in their Wyckoff positions, is key in determining the way in which they are impacted by QNEs. The first system to be described is PdH, which assumes a face-centered-cubic ($Fm\bar{3}m$) structure at ambient pressure ($1~\text{atm}=1.01325\times 10^{4}$~Pa, in our calculations). A hallmark of conventional, or electron-phonon mediated, superconductors is that their \tc\ decreases upon substitution of a lighter isotope by a heavier one, because lighter atoms experience higher frequency vibrations. In a conventional superconductor with one type of atom of mass $M$, it is expected that $T_\text{c} \propto M^{-\alpha}$, where $\alpha=0.5$ is the isotope coefficient. However, in PdH this coefficient is negative -- a consequence of QNEs~\cite{34} -- such that its measured \tc\ is $\sim$20\% lower than that of its heavier brethren, PdD~\cite{PhysRevB.29.4140,PhysRevB.12.117,PhysRevB.44.10377,PhysRevLett.39.574,PhysRevLett.34.144,schirber1974concentration,stritzker1972superconductivity}.  The next phase that we considered, AlH$_3$-$Pm\bar{3}n$, is one of the first high-pressure hydrides to be theoretically~\cite{PhysRevB.76.144114} and experimentally~\cite{PhysRevLett.100.045504} studied. Though calculations assuming classic nuclei predicted superconductivity, none could be measured down to 4~K~\cite{PhysRevLett.100.045504}, and this discrepancy was shown to be a consequence of QNEs~\cite{PhysRevB.82.104504,hou2021strong}. As already alluded to above, without the inclusion of QNEs LaH$_{10}$-$Fm\bar{3}m$ is predicted to be dynamically unstable~\cite{errea7} in a pressure range where it was synthesized and its superb superconducting properties were measured~\cite{somayazulu2019evidence,drozdov2019superconductivity}, becoming dynamically stable with classical nuclei only above $\sim$230 GPa. A hypothetical high-pressure phase of atomic hydrogen with $I4_1/amd$ symmetry computed to be the ground state above 577~GPa in the presence of QNEs with a \tc\ approaching 800~K at 2~TPa \cite{PhysRevB.84.144515,monacelli2023quantum,PhysRevLett.112.165501,PhysRevLett.114.105305} was also considered. As was a likely candidate of a synthesized hydride of platinum \cite{PhysRevB.83.214106} (PtH-$P6_3/mmc$), which was calculated to have a \tc\ of 12~K at 90~GPa~\cite{PhysRevB.83.214106,PhysRevLett.107.117002,PhysRevB.84.054543}, but whose superconductivity was shown to be suppressed upon the inclusion of QNEs~\cite{ErreaSSCHA}. In synthesized YH$_6$-$Im\bar{3}m$ the measured \tc\ (224~K at 166 GPa\cite{troyan2021anomalous} ) is lower than predictions that do not treat the nuclei quantum mechanically (264~K at 120 GPa\cite{heil2019superconductivity}). Turning to the experimentally synthesized H$_3$S phase~\cite{Drozdov:2015a}, we consider in this class the $Im\bar{3}m$ structure, which is predicted to become enthalpically preferred over the $R3m$ symmetry phase above 175~GPa~\cite{168,errea2016quantum} when treating ions classically, but QNEs extend its range of dynamic stability down to $\sim$100~GPa (though it may not be the thermodynamic minimum at this pressure). All these structures in this Type I class show a high symmetry in their bonding environment.
  
%

\begin{figure*}[t!]
\includegraphics[width=\textwidth]{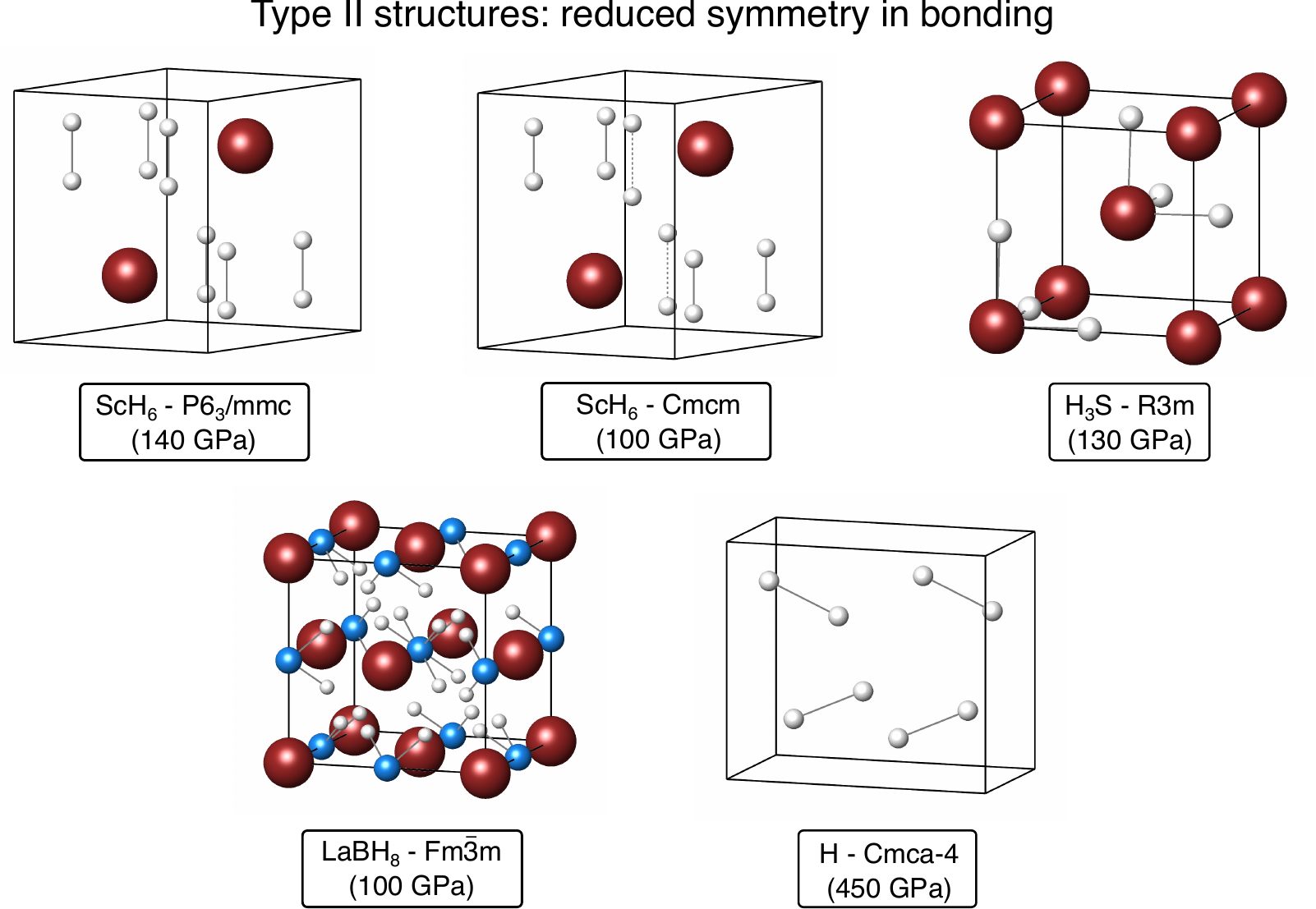}
\caption{\label{Fig:Type2}  Crystal structures of the Type II class of compounds. Locally, these phases exhibit a lower degree of symmetry in their bonding environments than the Type I compounds (Figure \ref{Fig:Type1}). Hydrogen atoms are white, B atoms in LaBH$_8$ are blue, and all other atom types are colored reddish-brown. The H-$Cmca$-4 notation corresponds to the primitive cell, while here the conventional cell is shown.}
\end{figure*}

The Type II family (Figure \ref{Fig:Type2}) includes two predicted ScH$_6$ phases with H$_2$ molecular units, which were thermodynamically preferred over the $Im\bar{3}m$ ScH$_6$ clathrate structure below 275~GPa~\cite{sch6prediction}. Diatomic hydrogen motifs comprise the hydrogenic sublattice in both, with the main difference between them being that the bond distances within the H$_2$ molecules in the higher pressure phase ($P6_3/mmc$) are the same, while in the lower pressure phase ($Cmcm$) they differ.  Moreover, a distorted H$_3$S structure, of $R3m$ symmetry, which DFT calculations predict to be enthalpically preferred over $Im\bar{3}m$ H$_3$S below 175~GPa was considered \cite{duan2014pressure,PhysRevB.91.180502}. In the rhombohedral distortion three H-S distances become shorter than the others, thereby breaking the octahedral symmetry about the sulfur atoms that is present in the cubic phase. This distortion is associated with a large measured drop in \tc, which was experimentally observed at 150~GPa signifying a small discrepancy between the predicted and measured pressure of the phase transition.\cite{Drozdov:2015a,einaga2016crystal,errea2015high} We also examined LaBH$_8$, the first ternary hydride proposed that can be derived from LaH$_{10}$ by removing the hydrogen atoms from the 8$c$ Wyckoff position, and adding boron atoms to the 4$a$ position.~\cite{Lucrezi,Di:2021bh,Liang} A number of compounds with this structure type, but varying the identity of the electropositive and $p$-block elements, have been predicted,~\cite{Zhang:2021,Di:2021bh} and superconductivity has been measured in isotypic LaBeH$_8$.~\cite{songStoichiometricTernarySuperhydride2023a} Finally, we considered the $Cmca$-4 phase of hydrogen~\cite{edwards1996layering}, composed of layered pairs of molecular hydrogen units in an ABAB hexagonal close packed stacking, which was predicted via static lattice calculations to be the most stable structure between 350-450 GPa~\cite{johnson2000structure}, or at 220~GPa and 300~K~\cite{PhysRevB.86.059902}, with a \tc\ of 242~K at 450~GPa~\cite{PhysRevLett.100.257001}. More recent investigations, accounting for QNEs, have proposed a refined phase diagram where a $Cmca$-12 phase is the most stable between 410-577~GPa, while also suggesting that $Cmca$-4 is not the ground state at any pressure.~\cite{monacelli2023quantum} The local bonding in all these Type II structures is less symmetric than those in the Type I class.

\subsection{Quantum nuclear effects on the structural and superconducting properties}
\begin{figure*}[t!]
\includegraphics[width=\textwidth]{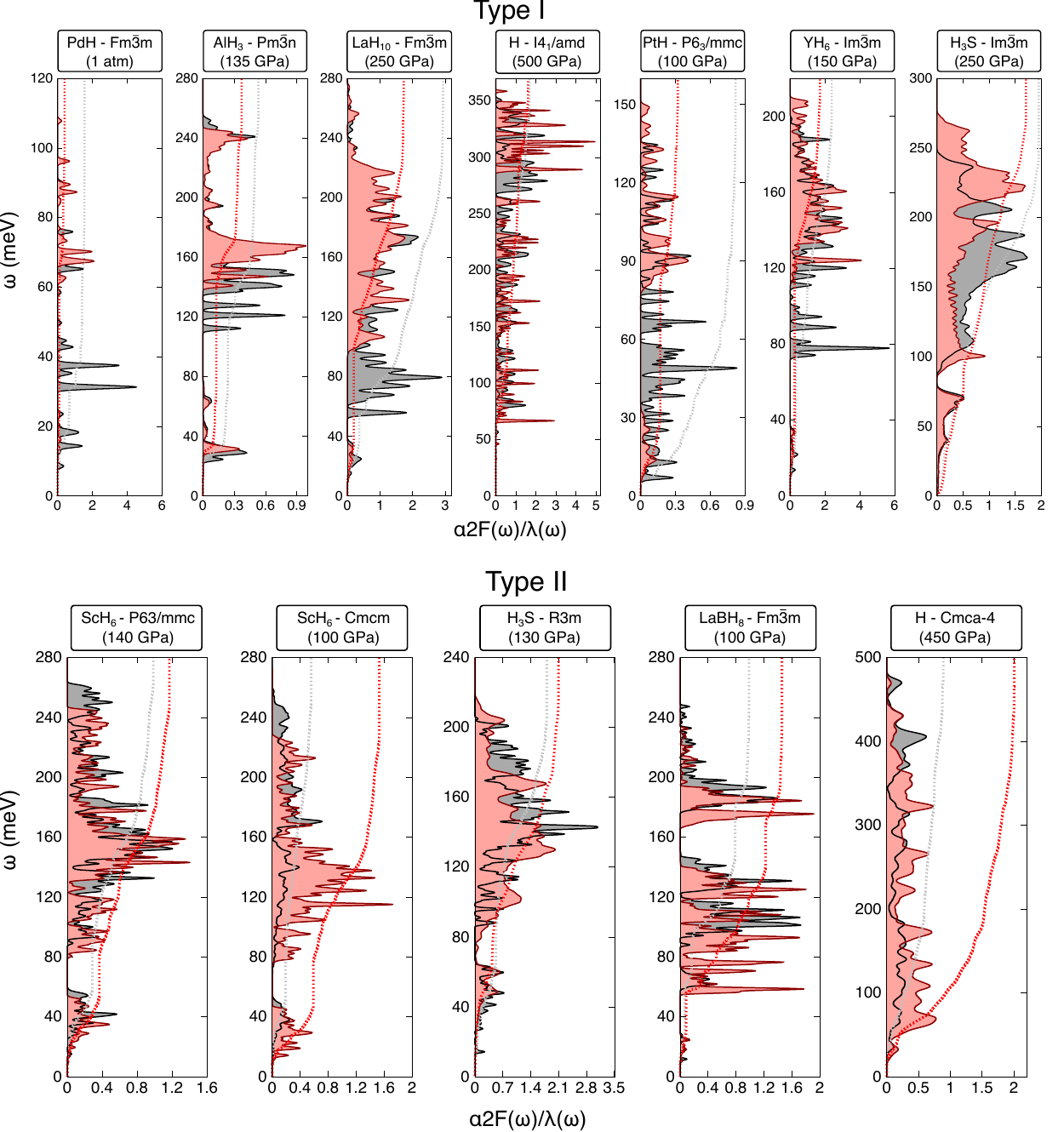}
\caption{\label{Fig:a2FA} The Eliashberg spectral function (shaded curve, $\alpha^2F(\omega)$), along with the integrated electron-phonon-coupling parameter (dashed line, $\lambda(\omega)$), for the Type I (upper panel) and Type II (lower panel) structures illustrated in Figure \ref{Fig:Type1} and Figure \ref{Fig:Type2}, respectively. Results for classical (quantum) nuclei are given in grey (red).}
\end{figure*}

Now that we have described the structural characteristics of the considered compounds, let us examine the properties that are related to their superconducting response. The key quantity for calculating the \tc\ in conventional superconductors is the Eliashberg spectral function,  $\alpha^2F(\omega)$, which can be understood as a phonon density of states, $F(\omega)$, weighted by the energy-dependent electron-phonon coupling, $\alpha(\omega)$. Though  $\alpha^2F(\omega)$ can, in principle, be measured (\emph{e.g.}\ via inelastic tunneling spectroscopy~\cite{PhysRevLett.114.047002}), it may be easier to calculate it via one of the various first-principles techniques that are available.  Figure \ref{Fig:a2FA} plots the variation of $\alpha^2F(\omega)$ as a function of frequency computed with classical nuclei and neglecting anharmonicity (grey), as compared to a SSCHA calculation incorporating quantum anharmonic effects on the structure and the phonons at zero temperature (red).  It is not uncommon for QNEs to expand the volume by a value that corresponds to $\sim$10~GPa~\cite{errea2020quantum}, but for clarity here the two calculations are performed at the same volume. This comparison allows us to estimate, at the same time, the impact of QNEs and anharmonicity on the phonon spectra and the electron-phonon matrix elements. If we were to compare the $\alpha^2F(\omega)$ functions computed at the same pressure, it would not be possible to conclude if the difference between them was a result of the different volumes, or because of QNEs.

Due to their high symmetry and absence of free parameters in the Wyckoff positions, relaxation incorporating QNEs does not affect the structural coordinates of most of the Type I compounds illustrated in Figure~\ref{Fig:Type1} (PdH-$Fm\bar{3}m$, AlH$_3$-$Pm\bar{3}n$, H-$I4_1/amd$, PtH-$P6_3/mmc$, YH$_6$-$Im\bar{3}m$, and H$_3$S-$Im\bar{3}m$). Because LaH$_{10}$-$Fm\bar{3}m$ has a single free parameter, which is related to the 32f Wyckoff site occupied by a hydrogen atom, QNEs can, in principle, alter this atomic position. However, after SSCHA relaxation we observed that the variation in this coordinate is negligible, consistent with the other phases belonging to this class of hydrides.

Comparison of the red and grey $\alpha^2F(\omega)$ curves in the top panel of Figure \ref{Fig:a2FA} reveals that QNEs tend to induce a blue shift of the phonon spectra in Type I compounds, and this effect is most pronounced for the lower optical branches. Because these modes contribute significantly to the EPC constant, and $\lambda \propto \omega^{-2}$, this phonon hardening tends to suppress $\lambda$ (Figure S2), and also \tc . At the pressures considered \tc\ decreases from 47~K to 5~K for PdH, 13.7~K to 3~K for AlH$_3$, 260~K to 220~K for LaH$_{10}$, 14.5~K to 0.4~K for PtH, 272~K to 247~K for YH$_6$, and  from 226~K to 190~K for H$_3$S (Table S3). The only exception to this trend is H-$I4_1/amd$, where QNEs produce a red shift of the acoustic phonon branches, but a blue shift of the optical phonon branches, resulting in a nearly negligible effect on both $\lambda$ and \tc\ (320~K versus 300~K), although when the volume expansion induced by QNEs is considered, the impact is a bit larger.~\cite{dangic_large_2024}

Let us now pivot now to the Type II structures shown in Figure \ref{Fig:Type2}, whose Eliashberg spectral functions are illustrated in the bottom panel of Figure \ref{Fig:a2FA}. Both of the ScH$_6$ phases considered contain dihydrogen molecules that interact weakly with one another, and more strongly with scandium via charge transfer from the electropositive element to the hydrogen, coupled with H$_2$ $\sigma \rightarrow$Sc d donation and Sc d $\rightarrow$H$_2$ $\sigma^*$ back-donation.~\cite{Zurek:2018b,Zurek:2020a,Zurek:2020g} These metal-H$_2$ interactions result in an elongation of the dihydrogen bond relative to what it would be in the molecular phase at the given pressure.~\cite{Zurek:2020g} In ScH$_6$-$P6_3/mmc$ at 140~GPa QNEs lengthen this bond even further, from 1.02 to 1.08~{\AA}, with a concomitant red shift of the phonon modes and an increase in \tc\ from 88~K to 99~K. When the nuclei are treated classically, the ground state of ScH$_6$ assumes $Cmcm$ symmetry at 100~GPa. This phase can be derived from the higher-pressure $P6_3/mmc$ structure by breaking the symmetry and decreasing the distance between two sets of hydrogen atoms to 0.95~\AA{}, and increasing the distance between another pair of H atoms to 1.28~\AA{}. Therefore, the classical ScH$_6$-$Cmcm$ lattice can be better described as [Sc$^{2+}$][2H$^-$]$\cdot$2H$_2$. However, quantum lattice fluctuations restore the broken symmetry, so all of the H-Sc bonds optimize to 1.08~{\AA} at 100~GPa within the SSCHA, resulting in a 5-fold increase in the \tc\ from 20~K to 108~K. 

A similar phenomenon occurs in H$_3$S-$R3m$, which experiments suggest is preferred over the higher symmetry $Im\bar{3}m$ structure below 150~GPa~\cite{einaga2016crystal}. At a volume corresponding to a classical pressure of 130~GPa, QNEs increase the pressure to $\sim$140~GPa. Moreover, they symmetrize the H-S bonds, which range from 1.46-1.66~{\AA} in the classic lattice, such that they all optimize to 1.56~{\AA} in the quantum lattice. The change in these bond lengths shifts the Eliashberg spectral function to lower frequencies, with a concomitant increase in the \tc\ from 175~K to 214~K. Though, in principle, a similar result would be expected for LaH$_{10}$ (at a pressure low enough so that the $Fm\bar{3}m$ phase is no longer the classical ground state~\cite{errea7}), we were unable to obtain a distorted lower-symmetry structure with a small enough unit cell to permit an \emph{ab initio} calculation of the electron-phonon interaction.

Because LaBH$_8$-$Fm\bar{3}m$ can be derived from LaH$_{10}$, its structure also possesses a free parameter for the hydrogen atoms situated at the 32f Wyckoff positions. Within LaBH$_8$ this parameter determines the B-H distance. Whereas in LaH$_{10}$-$Fm\bar{3}m$ the inclusion of QNEs does not have a notable impact on the positions of these hydrogen atoms, the behavior is different within LaBH$_8$-$Fm\bar{3}m$. Specifically, QNEs elongate and weaken the B-H bond from 1.366 to 1.385~{\AA}, with a concomitant softening of the frequency associated with its vibration. As a result,  the phonon spectra undergo a red shift and the \tc\ increases from 97~K to 143~K at 100~GPa. Additionally, QNEs tend to destabilize this structure due to the stretching of the B-H bond, rendering it unstable at 77~GPa, a much higher pressure than the 35 GPa  expected for classic nuclei~\cite{belli2022impact,DiCataldo:2021a}. This behavior is opposite to what has been found for H$_3$S~\cite{errea7} and LaH$_{10}$,\cite{errea2016quantum} where the pressure domain of dynamic stability is increased by QNEs. Finally, in the last Type II structure considered, a molecular H$_2$ phase with $Cmca$ symmetry and four hydrogen atoms in the primitive cell, QNEs elongate the H-H distances from 0.78 to 0.83~{\AA} at 400~GPa, weakening this bond with a concomitant red shift of the phonon modes, and a remarkable increase in the \tc\ from 109~K to 258~K.

\subsection{A descriptor for the impact of quantum nuclear effects on $T_\text{c}$}

\begin{figure*}[t]
\includegraphics[width=\textwidth]{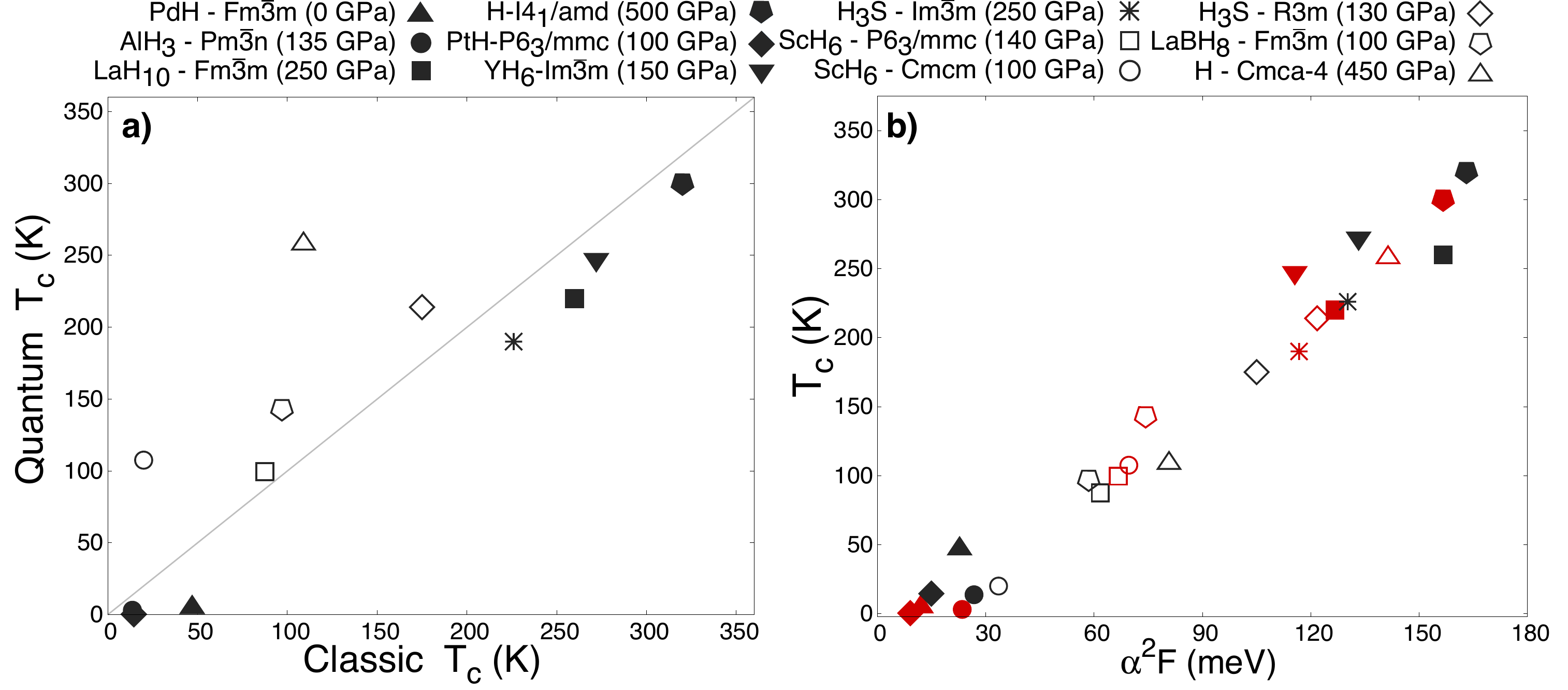}
\caption{\label{Fig:lambdaomega} (a) The superconducting critical temperature, \tc , computed for quantum nuclei versus classic nuclei for the Type 1 (Figure \ref{Fig:Type1}) and Type 2 (Figure \ref{Fig:Type2}) compounds studied herein. The open symbols (Type II) fall above the diagonal line and represent those compounds whose quantum \tc s are larger than their classic \tc s, while the quantum \tc s of the compounds denoted by the dark symbols (Type I) are lower than their classical \tc s. (b) The classical (black) and quantum (red) \tc s of the studied compounds as a function of the Eliashberg spectral function, $\alpha^2F(\omega)$. The legend above the plots provides the symbol associated with each compound.}
\end{figure*}

As described in the previous section, QNEs impact the geometries and structural features of the hydrogen-rich systems considered herein. Both these geometrical perturbations and anharmonicity can modify the frequencies associated with the phonon modes, which in turn affect the \tc . Because symmetry lowering (\emph{e.g.} Jahn-Teller or Peierls) distortions typically lead to the opening of gaps or pseudogaps, whereas symmetry raising transformations tend to increase the number of states that can participate in the EPC mechanism, QNE-induced structural changes could also vary the DOS at $E_\text{F}$, with a concomitant effect on the electron-phonon matrix elements and therefore on the resulting \tc.  For the systems studied here belonging to Type I, the practical absence of structural changes makes the DOS at $E_\text{F}$ insensitive to NQEs, therefore any variations in \tc\ were mostly a result of the changes in the phonon frequencies. The systems in Type II class, however, undergo structural changes that in general tend to increase the DOS at the Fermi level, making the electron-phonon coupling larger. This is evident for H-$Cmca$-4 and ScH$_6$-$Cmcm$, where the increased number of states at the Feemi level due to NQEs makes the value of the Eliashberg function much larger than in the classical nuclei case. Accordingly, in these cases the impact of NQEs is not only a shift of the phonon frequencies, but a more complex scenario in which the Fermi surface changes increase the coupling between the electrons and the ionic lattice. As highlighted by Figure \ref{Fig:lambdaomega}(a), QNEs typically decrease the \tc\ in Type I compounds, whereas for the Type II systems they tend to increase it, at times by a staggering factor of two (H-$Cmca$-4) or even four (ScH$_6$-$Cmcm$) when their impact on the Fermi surface is remarkable.

Interestingly, we have found that the integral of the Eliashberg spectral function, defined as $\alpha^2F = \int_0^{\infty} \alpha^2F(\omega) d\omega$, correlates very well with the \tc\ obtained for both classic harmonic and quantum anharmonic calculations (Figure \ref{Fig:lambdaomega}(b)). This finding indicates that $\alpha^2F$ itself is a robust descriptor for analysing how QNEs impact \tc. In fact, it correlates much better than the electron-phonon coupling constant, $\lambda$, and the average of the logarithmic phonon frequency, $\omega_\text{log}$, the prefactor in the Allen-Dynes modified McMillan semiempirical equation used to calculate \tc\ (Figure S2).

\begin{figure}[t]
\centering
\includegraphics[width=0.7\columnwidth]{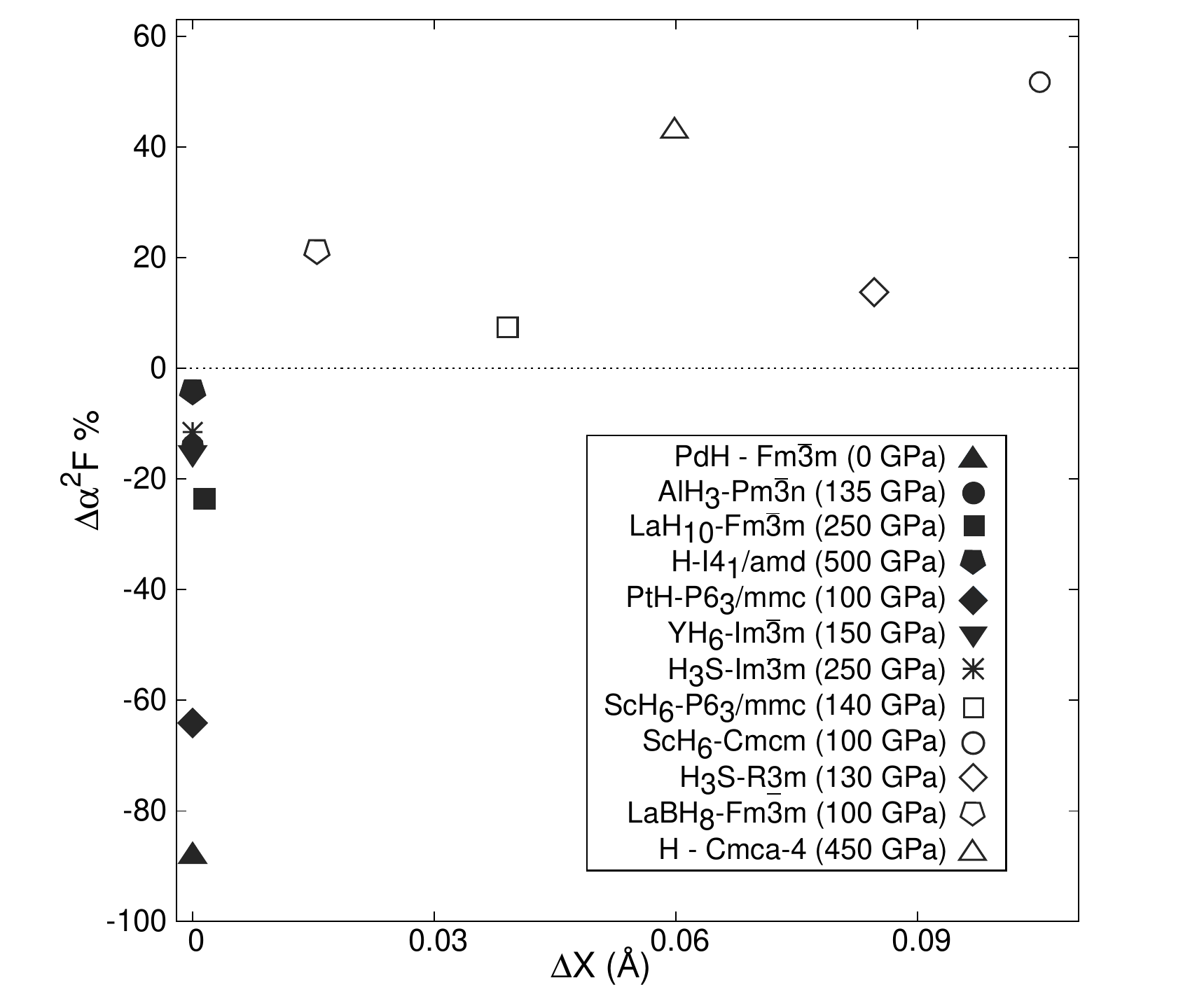}
\caption{\label{Fig:Distances} The amount by which the integral of the Eliashberg spectral function ($\alpha^2F = \int_0^{\infty} \alpha^2F(\omega) d\omega$) differs, given as a percentage, between calculations that treat the nuclei as quantum particles versus classical particles, as a function of the mean atomic displacement, $\Delta X$, defined in Equation \eqref{eq:displacement}. As shown in the legend, filled (open) symbols correspond to Type I (Type II) structures.}
\end{figure}

To better quantify the way in which QNEs affect each structure, we calculate the mean atomic displacements they impose as
\begin{equation}
    \Delta X = \sum_a^{N_A} \frac{\mid \boldsymbol{\mathcal{R}}^a_\text{q} - \textbf{R}^a_\text{c} \mid }{N_A},
    \label{eq:displacement}
\end{equation}
where $\boldsymbol{\mathcal{R}}^a_\text{q}$ is the quantum anharmonic position calculated for atom $a$, $\textbf{R}^a_\text{c}$ is the position of the same atom as obtained from the minimum of the BOES, and $N_A$ represents the number of atoms in the unit cell. Figure \ref{Fig:Distances} plots the percentage change in $\alpha^2F$ in going from the static to the quantum lattice as a function of $\Delta X$. It is clear that $\alpha^2F$ can change dramatically, even when $\Delta X$ does not. For the Type I structures, which are represented by full symbols, though the structure remains practically unaltered after the introduction of QNEs, a reduction for $\alpha^2F$ is observed, implying a decreased \tc . For the Type II systems, represented by the empty symbols, the exact opposite is observed:  the introduction of QNEs modifies the atomic parameters such that $\Delta X>0$ and \tc\ is increased. This analysis reveals that a comparison of the optimized geometries obtained assuming classic and quantum nuclei is sufficient to predict the way in which QNEs will affect the superconducting properties of hydrides. Our findings align with the analysis performed by Lucrezi \emph{et al}.\ on LaBH$_8$~\cite{lucrezi2023quantum}, who concluded that QNEs give rise to two general phenomena that affect superconducting properties: one driven by structural changes and another related to phonon-phonon interactions resulting from anharmonicity. Type I systems include those where the renormalization of the phonon-phonon interactions dominates over the structural perturbations, while the opposite phenomenon is characteristic of Type II compounds.

However, to obtain $\alpha^2F$ or $\Delta X$ a full geometry optimization and calculation of the phonon band structure and the EPC strength for both the classic and quantum lattice, the latter being particularly expensive even when accelerated using machine learning interatomic potentials, is still required. What we desire is a descriptor that predicts the way in which QNEs affect the geometric and superconducting properties of a structure that can be obtained knowing \emph{only} the geometry of the classic lattice. One might be tempted to conclude that crystals with higher symmetries are likely to be perturbed less by QNEs than ones with lower symmetries. However, such a rule of thumb would be incorrect, as the structurally LaH$_{10}$ (Type I) and LaBH$_8$ (Type II) compounds both assume the $Fm\bar{3}m$ spacegroup. Another example where spacegroups would not help differentiate between the two types of systems is PtH and ScH$_6$ at 140~GPa, which both adopt $P6_3/mmc$ symmetry, the former a member of the Type I family and the latter a member of Type II.

Since the total number of symmetry operations of the spacegroup and the total number of free parameters in its Wyckoff sites is insufficient to predict \emph{a priori} how QNEs will impact a structure, another metric must be found. Because our observations suggested that the local bonding environments about the atoms in a phase are key for predicting how it will be affected by QNEs, we first considered the vector weighted sum of the interatomic distances about an atom, and the Crystal Orbital Hamilton Population integrated to $E_\text{F}$ (iCOHP)~\cite{deringer2011crystal} between atom pairs, both summed over the unit cell, as descriptors (Table S6 and S7). However, neither approach demonstrated sufficient predictive power. Analyzing interatomic distances alone is insufficient because while one distance might correspond to a value typical of a single bond for one atom pair, for another atom pair it may be typical of a non-bonding or even a repulsive interaction. Though the iCOHP is sensitive to the bond strength, our tests showed that it was not able to predict which class LaH$_{10}$ and H-$Cmca$-4 would fall into (Table S6).

However, as shown below, an analysis of the local interatomic bonding patterns as calculated via the integral of the crystal orbital bonding index (iCOBI)~\cite{muller2021crystal} for pairs of atoms resulted in a predictive and simple-to-use descriptor. The iCOBI is a generalization of the bond index according to Wiberg and Mayer, which is directly related to the classic bond order, adapted to periodic systems. For a single bond, such as between two carbon atoms in diamond, an iCOBI of 1 would be expected, whereas in an ionic system, such as between the Na$^+$ and Cl$^-$ atoms in the rocksalt phase, iCOBI should be near 0. Previous computations yielded 0.95 and 0.09 for these cases~\cite{muller2021crystal}, in-line with our expectations. In a mixed covalent and ionic system, a value between 0 and 1 is therefore likely to be found. 

In the high pressure hydrides the interaction between hydrogen atoms is expected to yield iCOBIs that fall between 1 (for an H$_2$ molecule) and 0 (between hydridic hydrogen, H$^-$, or atomic hydrogen, H, and any other hydrogenic motif). When $p$-block elements form weak covalent bonds with the hydrogen atoms, such as in H$_3$S, the iCOBI is expected to be notably larger than 0, but smaller than 1. For example, in $Im\bar{3}m$ H$_3$S at 270~GPa the iCOBI between the S and H atoms was computed to be 0.34, indicating a bond order of roughly 1/3~\cite{Zurek:2021i}. While the interaction between electropositive elements and hydrogen is expected to be primarily ionic, we note that covalent (dative) interactions such as those arising from donation of orbitals from bonding states to a vacant d-orbital on the metal atom and back-bonding from an occupied metal d-orbital to antibonding hydrogenic states (\emph{e.g.} H$_2$ $\sigma \rightarrow$ metal d, and metal d $\rightarrow$ H$_2$ $\sigma^*$)~\cite{Zurek:2018b,Zurek:2020g,Zurek:2020a} are likely to be non-negligible resulting in non-zero iCOBIs. Importantly, we hypothesized that as long as the individual iCOBIs are not equal to naught, a vector sum of them about an atom can provide information about the symmetry of the atom's local bonding environment.

To test this hypothesis, we first introduced a vector quantity, $\vect{V}_x$, computed for each atom $x$. This vector is the sum of the iCOBI values for all of the interactions between an atom and its neighboring atoms $\alpha$ (iCOBI$(x,\alpha)$) weighted by a unit vector, $\vect{i}_{x\alpha}$, which denotes the direction of each interaction. This summation is performed over all neighbors for which the iCOBI values fall above a user-defined threshold, and divided by the number of interactions considered, $B_x$, as
\begin{equation}
     \vect{V}_{x} = \frac{1}{B_x}\sum_{\alpha=1}^{B_x}\text{iCOBI}(x,\alpha)\vect{i}_{x\alpha}.
     \label{eq:vec}
\end{equation}
The vector $\vect{V}_{x}$, is able to capture the asymmetry of the local bonding environment around the atom $x$ and quantify the effect of QNEs on the structure. However, $\vect{V}_{x}$ exhibits redundancy due to the symmetry of the system: atoms located on equivalent Wyckoff positions are expected to have identical magnitudes of $\vect{V}_{x}$. To simplify the analysis and reduce the number of variables, we investigated $S_x=\lvert \vect{V}_{x}\rvert$ for just one of the atoms belonging to the same Wyckoff position. We defined this magnitude as the symmetry index or parameter, denoted as ${S}_a$, where $a$ represents the group of atoms $x$ in the same Wyckoff position. 
The symmetry index is expected to be very small or zero for an atom that is in a completely symmetric bonding environment, whereas larger values of ${S}_a$ correlate with a greater degree of local bonding asymmetry around the atom. 

Is it possible to distinguish Type I structures from Type II based on the value of the symmetry parameters calculated for each symmetry inequivalent atom in the optimized classic lattice, $S_a^\text{c}$?  And, would the symmetry parameters for quantum nuclei, ${S}_{a}^\text{q}$, differ for these two classes of compounds? To answer these questions, ${S}_a^\text{c}$ and ${S}_a^\text{q}$ were calculated for each distinct atom in the systems illustrated in Figures \ref{Fig:Type1} and \ref{Fig:Type2}, and the results are provided in Table \ref{TABLE:Sa}. The analysis was performed using a cutoff threshold of 0.05 for the iCOBI in the calculation of $\vect{V}_x$ (Equation \ref{eq:vec}). This threshold was chosen to prioritize the most significant interactions and to minimize the inclusion of artifacts resulting from incomplete wavefunction projections onto localized orbitals. These interactions are reported in Tables S4 and S5.


\begin{table}[h!]

\caption{\label{TABLE:Sa} The symmetry inequivalent atoms and their Wyckoff positions in the Type I (Figure \ref{Fig:Type1}) and Type II (Figure \ref{Fig:Type2}) structures along with the corresponding symmetry index computed treating the nuclei as classical, ${S}_a^\text{c}$,  and quantum, ${S}_a^\text{q}$, objects. The Wyckoff parameters given in parenthesis for ScH$_6$ - $Cmcm$ correspond to the positions of these atoms in the higher symmetry ScH$_6$ - $P6_3/mmc$ phase.}
\begin{tabular}{|ccccc|}
\hline
\multicolumn{1}{|c|}{Structure}                      & \multicolumn{1}{c|}{Atom} & \multicolumn{1}{c|}{\begin{tabular}[c]{@{}c@{}}Wyckoff \\ letter\end{tabular}} & \multicolumn{1}{c|}{${S}_a^\text{c}$} & ${S}_a^\text{q}$ \\ \hline
\multicolumn{5}{|c|}{\multirow{2}{*}{Type I}}                        \\
\multicolumn{5}{|c|}{}                   \\ \hline
\multicolumn{1}{|c|}{\multirow{2}{*}{\begin{tabular}[c]{@{}c@{}}PdH - $Fm\bar{3}m$\\ (1 atm)\end{tabular}}} & \multicolumn{1}{c|}{Pd}  & \multicolumn{1}{c|}{4a}     & \multicolumn{1}{c|}{0 }  & 0    \\ \cline{2-5} 
\multicolumn{1}{|c|}{}   & \multicolumn{1}{c|}{H}   & \multicolumn{1}{c|}{4b}     & \multicolumn{1}{c|}{0}    & 0   \\ \hline
\multicolumn{1}{|c|}{\multirow{2}{*}{\begin{tabular}[c]{@{}c@{}}AlH$_3$ - $Pm\bar{3}m$\\ (135 GPa)\end{tabular}}}       & \multicolumn{1}{c|}{Al}  & \multicolumn{1}{c|}{2a}     & \multicolumn{1}{c|}{0}   & 0    \\ \cline{2-5} 
\multicolumn{1}{|c|}{}   & \multicolumn{1}{c|}{H}   & \multicolumn{1}{c|}{6c}     & \multicolumn{1}{c|}{0 }   & 0   \\ \hline
\multicolumn{1}{|c|}{\multirow{3}{*}{\begin{tabular}[c]{@{}c@{}}LaH$_{10}$ - $Fm\bar{3}m$\\ (250 GPa)\end{tabular}}}    & \multicolumn{1}{c|}{La}  & \multicolumn{1}{c|}{4b}     & \multicolumn{1}{c|}{0 }   & 0   \\ \cline{2-5} 
\multicolumn{1}{|c|}{}   & \multicolumn{1}{c|}{H}   & \multicolumn{1}{c|}{8c}     & \multicolumn{1}{c|}{0 }   & 0   \\ \cline{2-5} 
\multicolumn{1}{|c|}{}   & \multicolumn{1}{c|}{H}   & \multicolumn{1}{c|}{32f}    & \multicolumn{1}{c|}{0.009}  & 0.009 \\ \hline
\multicolumn{1}{|c|}{\begin{tabular}[c]{@{}c@{}}H - $I4_1/amd$\\ (500 GPa)\end{tabular}}                          & \multicolumn{1}{c|}{H}   & \multicolumn{1}{c|}{4b}     & \multicolumn{1}{c|}{0 }  & 0    \\ \hline
\multicolumn{1}{|c|}{\multirow{2}{*}{\begin{tabular}[c]{@{}c@{}}PtH - $P6_3/mmc$\\ (100 GPa)\end{tabular}}}           & \multicolumn{1}{c|}{Pt}  & \multicolumn{1}{c|}{2d}     & \multicolumn{1}{c|}{0 }   & 0   \\ \cline{2-5} 
\multicolumn{1}{|c|}{}   & \multicolumn{1}{c|}{H}   & \multicolumn{1}{c|}{2a}     & \multicolumn{1}{c|}{0 }   & 0   \\ \hline
\multicolumn{1}{|c|}{\multirow{2}{*}{\begin{tabular}[c]{@{}c@{}}YH$_6$ - $Im\bar{3}m$\\ (150 GPa)\end{tabular}}}        & \multicolumn{1}{c|}{Y}   & \multicolumn{1}{c|}{2a}     & \multicolumn{1}{c|}{0}    & 0   \\ \cline{2-5} 
\multicolumn{1}{|c|}{}   & \multicolumn{1}{c|}{H}   & \multicolumn{1}{c|}{12d}    & \multicolumn{1}{c|}{0 }   & 0   \\ \hline
\multicolumn{1}{|c|}{\multirow{2}{*}{\begin{tabular}[c]{@{}c@{}}H$_3$S - $Im\bar{3}m$\\ (250 GPa)\end{tabular}}} & \multicolumn{1}{c|}{S}   & \multicolumn{1}{c|}{2a}     & \multicolumn{1}{c|}{0 }  & 0    \\ \cline{2-5} 
\multicolumn{1}{|c|}{}   & \multicolumn{1}{c|}{H}   & \multicolumn{1}{c|}{6b}     & \multicolumn{1}{c|}{0 }   & 0   \\ \hline
\multicolumn{5}{|c|}{\multirow{2}{*}{Type II}}                       \\
\multicolumn{5}{|c|}{}                   \\ \hline
\multicolumn{1}{|c|}{\multirow{2}{*}{\begin{tabular}[c]{@{}c@{}}ScH$_6$ - $P6_3/mmc$\\ (140 GPa)\end{tabular}}}   & \multicolumn{1}{c|}{Sc}  & \multicolumn{1}{c|}{2d}     & \multicolumn{1}{c|}{0}    & 0   \\ \cline{2-5} 
\multicolumn{1}{|c|}{}   & \multicolumn{1}{c|}{H}   & \multicolumn{1}{c|}{12k}    & \multicolumn{1}{c|}{0.056} & 0.054 \\ \hline
\multicolumn{1}{|c|}{\multirow{3}{*}{\begin{tabular}[c]{@{}c@{}}ScH$_6$ - $Cmcm$\\ (100 GPa)\end{tabular}}}       & \multicolumn{1}{c|}{Sc}  & \multicolumn{1}{c|}{4c}     & \multicolumn{1}{c|}{0.012}  & 0 (2d)\\ \cline{2-5} 
\multicolumn{1}{|c|}{}   & \multicolumn{1}{c|}{H}   & \multicolumn{1}{c|}{16h}    & \multicolumn{1}{c|}{0.040} & 0.055 (12k) \\ \cline{2-5} 
\multicolumn{1}{|c|}{}   & \multicolumn{1}{c|}{H}   & \multicolumn{1}{c|}{8f}     & \multicolumn{1}{c|}{0.082}  & 0.055 (12k) \\ \hline
\multicolumn{1}{|c|}{\multirow{2}{*}{\begin{tabular}[c]{@{}c@{}}H$_3$S - $R3m$\\ (130 GPa)\end{tabular}}}         & \multicolumn{1}{c|}{S}   & \multicolumn{1}{c|}{3a}     & \multicolumn{1}{c|}{0.076}  & 0 \\ \cline{2-5} 
\multicolumn{1}{|c|}{}   & \multicolumn{1}{c|}{H}   & \multicolumn{1}{c|}{9b}     & \multicolumn{1}{c|}{0.116} & 0 \\ \hline
\multicolumn{1}{|c|}{\multirow{3}{*}{\begin{tabular}[c]{@{}c@{}}LaBH$_8$ - $Fm\bar{3}m$\\ (100 GPa)\end{tabular}}}      & \multicolumn{1}{c|}{La}  & \multicolumn{1}{c|}{4b}     & \multicolumn{1}{c|}{0 }     & 0 \\ \cline{2-5} 
\multicolumn{1}{|c|}{}   & \multicolumn{1}{c|}{B}   & \multicolumn{1}{c|}{4a}     & \multicolumn{1}{c|}{0}     & 0  \\ \cline{2-5} 
\multicolumn{1}{|c|}{}   & \multicolumn{1}{c|}{H}   & \multicolumn{1}{c|}{32f}    & \multicolumn{1}{c|}{0.062} & 0.063 \\ \hline
\multicolumn{1}{|c|}{\begin{tabular}[c]{@{}c@{}}H - $Cmca$-4\\ (450 GPa)\end{tabular}}                       & \multicolumn{1}{c|}{H}   & \multicolumn{1}{c|}{8f}     & \multicolumn{1}{c|}{0.022} & 0.014 \\ \hline
\end{tabular}
\end{table}

Examination of Table \ref{TABLE:Sa} clearly shows that for Type I structures both the classic and quantum asymmetry parameters are the same and are either zero, or very nearly zero, for each symmetry inequivalent atom.  In fact, the only non-zero value that was calculated was for the hydrogen atoms at the 32f Wyckoff position in LaH$_{10}$, which has a free parameter. In the lattice optimized with QNEs, this value does not change, in-line with the above discussion where it was shown that the quantum nature of the hydrogen atoms and the resulting anharmonicity had a negligible effect on the optimized atomic positions. The value of $S_a^\text{c}$ for this Wyckoff position, 0.009,  is an order of magnitude smaller than the non-zero values computed for any of the Type II structures. Thus, we suggest that the classical lattice symmetry index, ${S}_a^\text{c}$, should be less than or equal to 0.01 for every atom within hydrides that belong to the Type I family, that is those where QNEs do not alter the structure, but where the phonon modes are shifted to higher frequencies because of the phonon-phonon interactions. These combined effects reduce $\lambda$, as well as $\alpha^2F$, leading to a concomitant decrease in \tc\ in Type I systems.  Our analysis rationalizes why LaH$_{10}$ belongs to the Type I family: despite the intrinsic freedom of the 32f Wyckoff site, the local bonding environment about these atoms is sufficiently symmetric to ensure that the atomic positions are not strongly altered by QNEs. 

For Type II structures, on the other hand, at least one of the symmetry inequivalent atoms possesses a significant $S_a^\text{c}$. Interestingly, the symmetry parameters for all but two of the heavy atoms are zero, in-line with the finding that QNEs do not perturb their positions during the relaxation, such that $S_a^\text{q}$ is also zero. The two exceptions are the Sc atoms within  ScH$_6$-$Cmcm$ and the S atoms within H$_3$S-$R3m$, where the classical symmetry indices of 0.022 and 0.076, respectively, decreased to zero during structural relaxation with the SSCHA, resulting in a symmetric environment about these atoms. In-line with the expectation that the lighter hydrogen atoms should have a higher proclivity to be perturbed by QNEs than the heavy atoms, they had significantly larger $S_a^\text{c}$ values (0.046 - 0.852), which all decreased substantially when their lattices were optimized with the SSCHA. 

These results highlight that the iCOBI based symmetry parameter is an excellent descriptor for identifying the atoms that will be most affected by QNEs in hydrogen rich materials including those that are stable at low and at high pressures. Moreover, we can conclude that compounds with at least one large $S_a^\text{c}$ value tend to increase the symmetry in their local bonding environment upon structural relaxation with QNEs, in particular around the hydrogen atoms, shifting the phonon modes to lower frequencies. These effects serve to increase $\lambda$, as well as $\alpha^2F$, and the concomitant \tc , in stark contrast to what is observed in Type I systems.

\section{Conclusions}

This work investigates how quantum nuclear effects (QNEs) and the anharmonicity they can induce alter the structural and superconducting properties of hydrogen-based superconductors. We introduce a symmetry bonding index or parameter, $S_a$, for each symmetry inequivalent atom, which is based upon a vector weighted sum of the integrated crystal orbital bonding indices (iCOBIs) between this atom and others falling above a user-defined threshold (here chosen as iCOBI=0.05). An $S_a$ value of zero corresponds to an atom in a nearly perfect symmetric bonding configuration, whereas larger values of $S_a$ are associated with non-negligible deviations from ideal symmetry. This index can be computed for lattices of classical nuclei optimized within the Born-Oppenheimer approximation, $S_a^\text{c}$, as well as relaxations that treat the nuclei quantum mechanically, $S_a^\text{q}$. Importantly, we find that the classical results are sufficient to predict the way in which QNEs will impact the properties of both ambient pressure and high pressure hydride superconductors. Thus, $S_a^\text{c}$, which can be calculated knowing only the positions of the classical lattice, is a robust and quick-to-compute descriptor that can predict how QNEs and anharmonicity affect the geometry of a compound and its superconducting behavior.

Two families of structures were identified and analyzed. In Type I systems $S_a^\text{c}$ was calculated to be zero for most atom types, and we suggest that it should not exceed 0.01 for any atom belonging to phases in this class. QNEs do not perturb the geometries of Type I compounds, but anharmonicity lowers  the pressure at which they are predicted to become dynamically unstable, with a consequent blue shift of their phonon spectra, which is stronger for the lower optical branches. This perturbation of the phonon modes results in a decrease of the \tc . In Type II systems, on the other hand, $S_a^\text{c}$ is large for at least one of the symmetry inequivalent atoms, meaning that these phases have a lower degree of local bonding symmetry. Structural relaxation with QNEs attempts to restore the local symmetry, often (but not always) resulting in smaller $S_a^\text{q}$ values. This geometric perturbation of the quantum lattice destabilizes the structure, softens the phonon modes, and enhances \tc , sometimes by an impressive factor of 2-4 when QNEs increase the density of states at the Fermi level. We expect the introduced classical asymmetry parameter will become a powerful tool in the \emph{a priori} prediction of the effect of QNEs on the geometries, phonon modes and critical temperatures of hydride superconductors, and its quantum counterpart will be used to understand the magnitude of these effects for specific lattices and atom types.


\section*{Supporting Information}
The Supporting Information is available free of charge on the ACS Publication website. It includes: the electronic configurations of the PAW potentials used in the LOBSTER calculations, the unit cell parameters and atomic positions for all of the structures optimized in the Born-Oppenheimer approximations, tables listing the classical and quantum $\lambda$, $\omega_\text{log}$ and \tc\ , plots of the quantum versus the classical $\lambda$ and $\omega_\text{log}$, the values of the most relevant bonds used in the iCOBI analysis, and the data for the analysis performed with iCOHPs and interatomic distances.

\section*{Acknowledgments}
Funding for this research is provided by the National Science Foundation, under award
DMR-2136038 and the European Research Council (ERC) under the European Union's Horizon 2020 research and innovation program (Grant Agreement No. 802533). I.E. also acknowledges financial support from the Department of Education, Universities and Research of the Eusko Jaurlaritza, and the University of the Basque Country UPV/EHU (Grant No. IT1527-22); and the Spanish Ministerio de Ciencia e Innovación (Grant No. PID2022-142861NA-I00). Calculations were performed at the Center for Computational Research at SUNY Buffalo
(http://hdl.handle.net/10477/79221).  


\providecommand{\latin}[1]{#1}
\makeatletter
\providecommand{\doi}
  {\begingroup\let\do\@makeother\dospecials
  \catcode`\{=1 \catcode`\}=2 \doi@aux}
\providecommand{\doi@aux}[1]{\endgroup\texttt{#1}}
\makeatother
\providecommand*\mcitethebibliography{\thebibliography}
\csname @ifundefined\endcsname{endmcitethebibliography}
  {\let\endmcitethebibliography\endthebibliography}{}

\newpage

\textbf{Table of Contents Graphic}

\begin{figure*}
\begin{center}
\includegraphics[width=8.25cm]{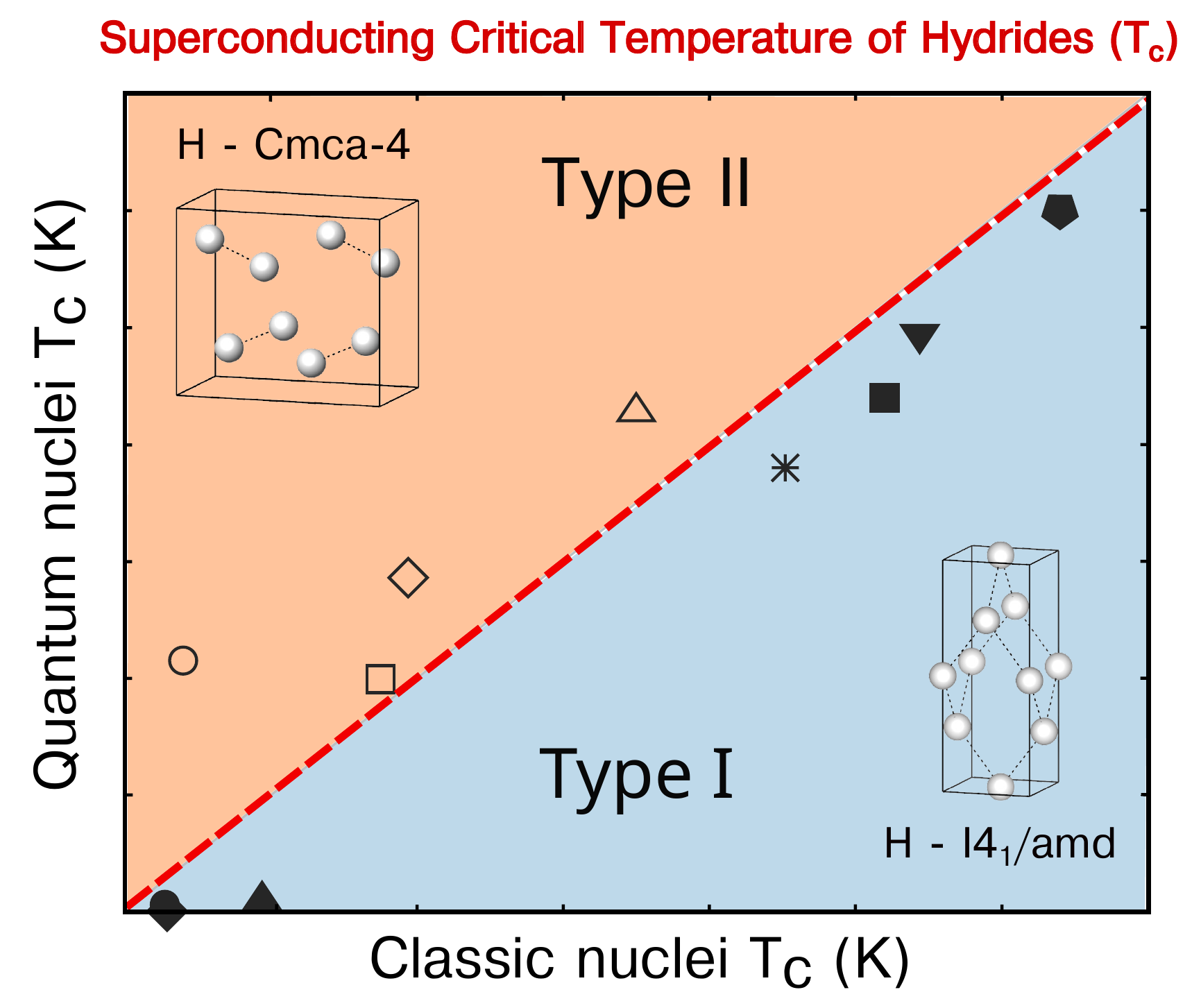}
\end{center}
\end{figure*}

\pagebreak


 


\begin{center}
\textbf{\Large Supplemental Materials: \\ A chemical bonding based descriptor for predicting the impact of quantum nuclear and anharmonic effects on hydrogen-based superconductors}
\end{center}

\setcounter{equation}{0}
\setcounter{figure}{0}
\setcounter{table}{0}
\setcounter{page}{1}
\makeatletter
\renewcommand{\theequation}{S\arabic{equation}}
\renewcommand{\thefigure}{S\arabic{figure}}
\renewcommand{\bibnumfmt}[1]{[S#1]}
\renewcommand{\citenumfont}[1]{S#1}

\section{Projector Augmented Wave (PAW) Potentials}

\begin{table}[]
\caption{The table reports the file name and valence electronic configurations for the PAW potentials used in the \textsc{VASP} calculations.}
\begin{tabular}{lll}
Element & File Name  & Valence Configuration       \\ \hline\hline
H    &  PAW\_PBE H   & 1s$^1$                   \\
B    &  PAW\_PBE B   & 2s$^2$2p$^1$             \\
Al   &  PAW\_PBE Al  & 3s$^2$3p$^1$             \\
S    & PAW\_PBE S    & 3s$^2$3p$^4$             \\
Sc   & PAW\_PBE Sc   & 3d$^2$4s$^1$             \\
Y    & PAW\_PBE Y\_sv & 4s$^2$5s$^2$4p$^6$4d$^2$ \\
Pd   & PAW\_PBE Pd   & 4d$^9$5s$^1$             \\
Pt   & PAW\_PBE Pt   & 5d$^9$6s$^1$             \\
La   & PAW\_PBE La   & 5s$^2$6s$^2$5p$^6$5d$^1$ \\ \hline\hline
\end{tabular}
\end{table}

\section{Phonons and Eliashberg Function of ScH$_6$-$Cmcm$}
\begin{figure*}[h]
\includegraphics[width=0.85\textwidth]{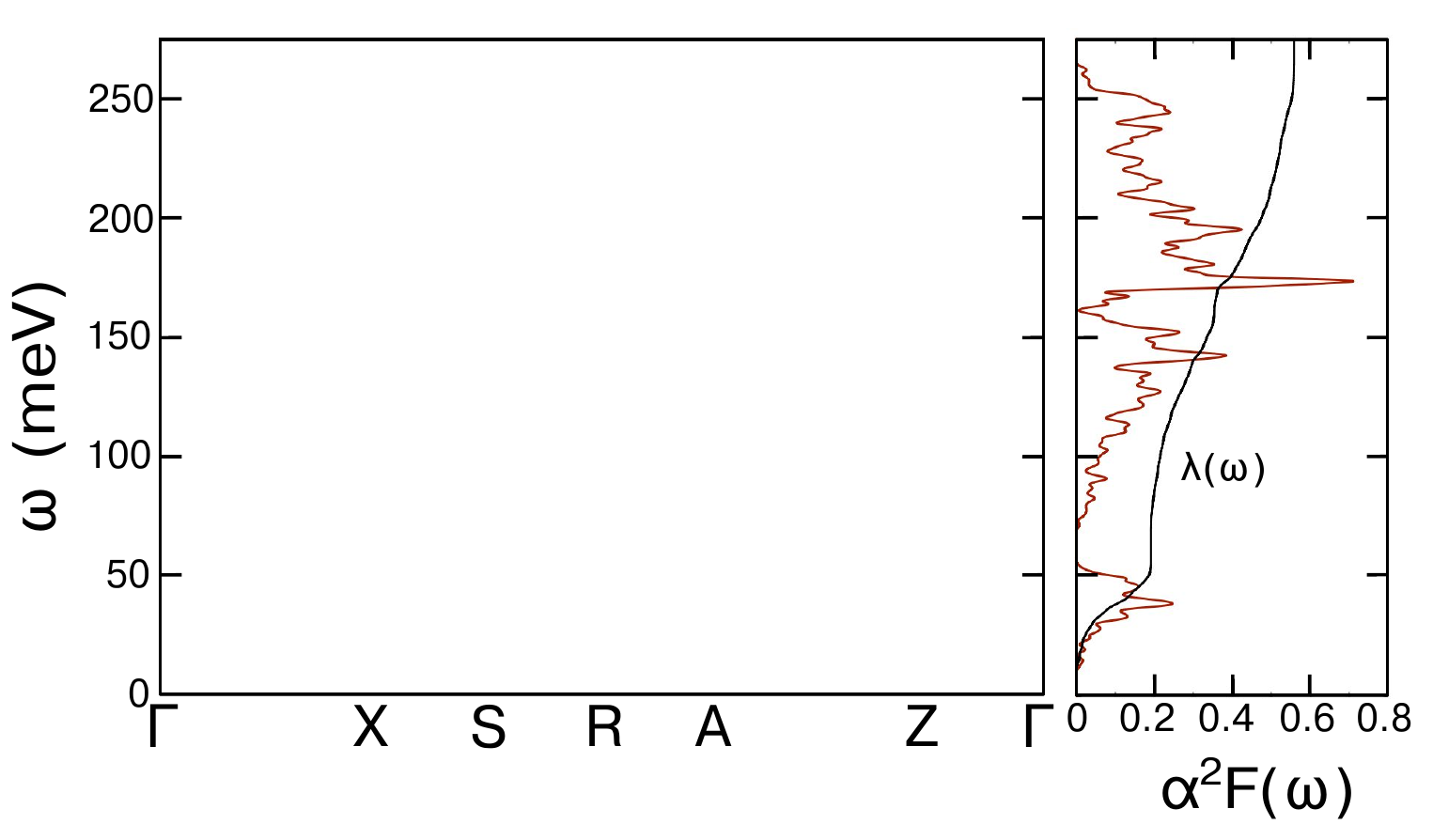}
\caption{\label{Fig:ScH6Cmcm} The classic phonon spectra for the ScH$_6$ - $Cmcm$ phase at 100~GPa (left), and the variation of the Eliashberg spectral function, $\alpha^2F(\omega)$, along with the integral of the electron phonon coupling parameter , $\lambda(\omega)$ (right), with the frequency.}
\end{figure*}

\newpage
\section{Structural Parameters}

\begin{table}[h]
\caption{The cell parameters and atomic positions for the considered compounds obtained when the nuclei were treated as classic particles.}
\begin{tabular}{cllll}
\hline \hline
Structure      & Parameters                    & Atom & \begin{tabular}[c]{@{}l@{}}Wyckoff Site \end{tabular} & Coordinate       \\ \hline
\multirow{2}{*}{\begin{tabular}[c]{@{}c@{}}PdH - $Fm\bar{3}m$\\ (1 atm)\end{tabular}}          & \multirow{2}{*}{\textit{a} = 4.132 \AA}                     & Pd   & 4a                & 0.00  0.00  0.00 \\
               &           & H    & 4b                & 0.50  0.50  0.50 \\ \hline
\multirow{2}{*}{\begin{tabular}[c]{@{}c@{}}AlH$_3$ - $Pm\bar{3}n$\\ (135 GPa)\end{tabular}}       & \multirow{2}{*}{\textit{a} = 3.003 \AA}                     & Al   & 2a                & 0.00  0.00  0.00 \\
               &           & H    & 6c                & 0.50  0.25  0.00 \\ \hline
\multirow{3}{*}{\begin{tabular}[c]{@{}c@{}}LaH$_{10}$ - $Fm\bar{3}m$\\ (250 GPa)\end{tabular}} & \multirow{3}{*}{\textit{a} = 3.424 \AA}                     & La   & 4b                & 0.50  0.50  0.50 \\
               &           & H    & 8c                & 0.25  0.25  0.25 \\
               &           & H    & 32f               & 0.10  0.10  0.10 \\ \hline
\begin{tabular}[c]{@{}c@{}}H - $I4_1/amd$\\ (500 GPa)\end{tabular}         & \begin{tabular}[c]{@{}l@{}}\textit{a} = 1.120 \AA\\ \textit{c} = 3.110 \AA\end{tabular}   & H    & 4b                & 0.50  0.50  0.00 \\ \hline
\multirow{3}{*}{\begin{tabular}[c]{@{}c@{}}PtH - $P6_3/mmmc$\\ (100 GPa)\end{tabular}}         & \multirow{3}{*}{\begin{tabular}[c]{@{}l@{}}\textit{a} = 2.710 \AA\\ \textit{c} = 4.576 \AA\\ $\gamma$ = 120$^{\circ}$\end{tabular}}                & Pt   & 2d                & 0.33 0.67  0.75  \\
               &           & H    & 2a                & 0.00  0.00  0.00 \\
               &      &  \\\hline               
\multirow{2}{*}{\begin{tabular}[c]{@{}c@{}}YH$_{6}$ - $Im\bar{3}m$\\ (150 GPa)\end{tabular}}   & \multirow{2}{*}{\textit{a} = 4.103 \AA}                     & Y    & 2a                & 0.00  0.00  0.00 \\
               &           & H    & 12d               & 0.25  0.00  0.50 \\ \hline
\multirow{2}{*}{\begin{tabular}[c]{@{}c@{}}H$_{3}$S - $Im\bar{3}m$\\ (250 GPa)\end{tabular}}   & \multirow{2}{*}{\textit{a} = 3.007 \AA}                     & S    & 2a                & 0.00  0.00  0.00 \\
               &           & H    & 6b               & 0.50  0.50  0.00 \\ \hline
\multirow{3}{*}{\begin{tabular}[c]{@{}c@{}}ScH$_{6}$ - $P6_3/mmc$\\ (140 GPa)\end{tabular}}    & \multirow{3}{*}{\begin{tabular}[c]{@{}l@{}}\textit{a} = 3.406 \AA\\ \textit{c} = 4.255 \AA\\ $\gamma$ = 120$^{\circ}$\end{tabular}}                & Sc   & 2d                &  0.67 0.33  0.25 \\
               &           & H    & 12k               & \multicolumn{1}{c}{0.67  0.83  0.13} \\ 
               &  & \\\hline
      
\multirow{3}{*}{\begin{tabular}[c]{@{}c@{}}ScH$_{6}$ - $Cmcm$\\ (100 GPa)\end{tabular}}        & \multirow{3}{*}{\begin{tabular}[c]{@{}l@{}}\textit{a} = 3.434\AA\\ \textit{b} =  6.215 \AA\\ \textit{c} = 4.433 \AA\\\end{tabular}} & Sc   & 4c                & 0.00  0.33  0.75 \\
               &           & H    & 16h               & 0.75    0.43      0.36      \\
               &           & H    & 8f                & 0.83  0.66  0.89 \\ \hline
\multirow{2}{*}{\begin{tabular}[c]{@{}c@{}}H$_{3}$S - $R3m$\\ (130 GPa)\end{tabular}}          & \multirow{2}{*}{\textit{a} = 3.116 \AA}                     & S    & 3a                & 0.00  0.00  0.00 \\
               &           & H    & 9b                & 0.53  0.01  0.01 \\ \hline
\multirow{3}{*}{\begin{tabular}[c]{@{}c@{}}LaBH$_{8}$ - $Fm\bar{3}m$\\ (200 GPa)\end{tabular}} & \multirow{3}{*}{\textit{a} = 5.577 \AA}                     & La   & 4b                & 0.50  0.50  0.50 \\
               &           & B    & 4a                & 0.00  0.00  0.00 \\
               &           & H    & 32f               & 0.15  0.15  0.15 \\ \hline
\begin{tabular}[c]{@{}c@{}}H - $Cmca$-4\\ (450 GPa)\end{tabular}           & \begin{tabular}[c]{@{}l@{}}\textit{a} = 1.534 \AA\\ \textit{b} = 2.674 \AA\\ \textit{c} = 2.360 \AA\end{tabular}        & H    & 8f                & 0.00 0.37  0.43  \\ \hline \hline
\end{tabular}
\end{table}

\newpage
\section{Superconductivity}

\begin{figure*}[h]
\includegraphics[width=\textwidth]{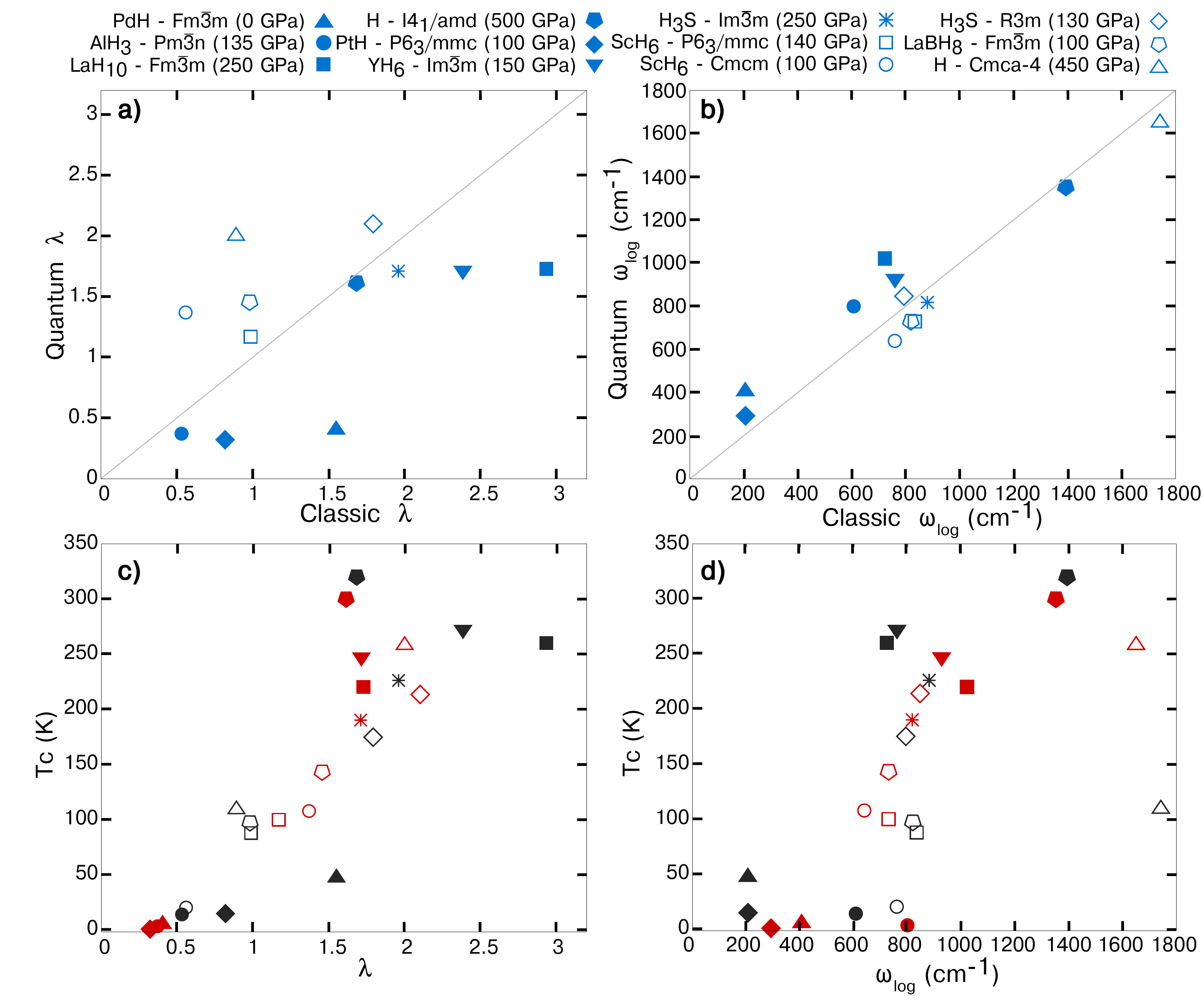}
\caption{A comparison of (a) the electron phonon coupling parameter, $\lambda$, and (b) the logarithmic average phonon frequency, $\omega_{\mathrm{log}}$, between the classical and quantum structures. The line drawn in (a) and (b) corresponds to $y=x$. (c) The classic (black) and quantum (red) values of \tc\ as a function of  $\lambda$. (d) The classic (black) and quantum (red) \tc\ as a function of $\omega_{\mathrm{log}}$. }

\end{figure*}

\begin{table}[]
\caption{The electron phonon coupling ($\lambda$), logarithmic average phonon frequency ($\omega_{\mathrm{log}}$), and superconducting critical temperature ($T_{\mathrm{c}}$), for the classical and quantum structures divided into Type I (top) and Type II (bottom) families.
}
\begin{tabular}{cccc|ccc}
\hline \hline
 \multirow{3}{*}{Structure}                  & \multicolumn{3}{|c|}{Classic}   & \multicolumn{3}{c}{Quantum}     \\
\multicolumn{1}{c|}{}      & $\lambda$            & \begin{tabular}[c]{@{}c@{}}$\omega_{\mathrm{log}}$\\  (cm$^{-1}$)\end{tabular} & \begin{tabular}[c]{@{}c@{}}$T_{\mathrm{c}}$ \\ (K)\end{tabular} & $\lambda$            & \begin{tabular}[c]{@{}c@{}}$\omega_{\mathrm{log}}$ \\ (cm$^{-1}$)\end{tabular} & \begin{tabular}[c]{@{}c@{}}$T_{\mathrm{c}}$\\  (K)\end{tabular} \\ \hline
\multicolumn{1}{l|}{}     & \multicolumn{1}{l}{} & \multicolumn{1}{l}{}    & \multicolumn{1}{l|}{}                & \multicolumn{1}{l}{} & \multicolumn{1}{l}{}    & \multicolumn{1}{l}{}                 \\
\multicolumn{1}{c|}{\begin{tabular}[c]{@{}c@{}}PdH - $Fm\bar{3}m$\\ (1 atm)\end{tabular}}           & 1.55                 & 205  & 47        & 0.40                & 405  & 5         \\
\multicolumn{1}{l|}{}     & \multicolumn{1}{l}{} & \multicolumn{1}{l}{}    & \multicolumn{1}{l|}{}                & \multicolumn{1}{l}{} & \multicolumn{1}{l}{}    & \multicolumn{1}{l}{}                 \\
\multicolumn{1}{c|}{\begin{tabular}[c]{@{}c@{}}AlH$_3$ - $Pm\bar{3}n$\\  (135 GPa)\end{tabular}}    & 0.53                 & 607  & 13.7      & 0.37                & 799  & 3.08      \\
\multicolumn{1}{l|}{}     & \multicolumn{1}{l}{} & \multicolumn{1}{l}{}    & \multicolumn{1}{l|}{}                & \multicolumn{1}{l}{} & \multicolumn{1}{l}{}    & \multicolumn{1}{l}{}                 \\
\multicolumn{1}{c|}{\begin{tabular}[c]{@{}c@{}}LaH$_{10}$ - $Fm\bar{3}m$\\  (250 GPa)\end{tabular}} & 2.93                 & 722  & 260       & 1.72                 & 1021 & 220       \\
\multicolumn{1}{l|}{}     & \multicolumn{1}{l}{} & \multicolumn{1}{l}{}    & \multicolumn{1}{l|}{}                & \multicolumn{1}{l}{} & \multicolumn{1}{l}{}    & \multicolumn{1}{l}{}                 \\
\multicolumn{1}{c|}{\begin{tabular}[c]{@{}c@{}}H - $I4_1/amd$ \\ (500 GPa)\end{tabular}}            & 1.68                 & 1393 & 320       & 1.61                 & 1351 & 300       \\
\multicolumn{1}{l|}{}     & \multicolumn{1}{l}{} & \multicolumn{1}{l}{}    & \multicolumn{1}{l|}{}                & \multicolumn{1}{l}{} & \multicolumn{1}{l}{}    & \multicolumn{1}{l}{}                 \\
\multicolumn{1}{c|}{\begin{tabular}[c]{@{}c@{}}PtH - $P6_3/mmc$\\ (100 GPa)\end{tabular}}           & 0.82                & 205  & 14.5      & 0.32                & 291  & 0.378    \\
\multicolumn{1}{l|}{}     & \multicolumn{1}{l}{} & \multicolumn{1}{l}{}    & \multicolumn{1}{l|}{}                & \multicolumn{1}{l}{} & \multicolumn{1}{l}{}    & \multicolumn{1}{l}{}                 \\
\multicolumn{1}{c|}{\begin{tabular}[c]{@{}c@{}}YH$_6$ - $Im\bar{3}m$\\  (150 GPa)\end{tabular}}     & 2.38                 & 760  & 272       & 1.71                 & 926  & 247       \\ 
\multicolumn{1}{l|}{}     & \multicolumn{1}{l}{} & \multicolumn{1}{l}{}    & \multicolumn{1}{l|}{}                & \multicolumn{1}{l}{} & \multicolumn{1}{l}{}    & \multicolumn{1}{l}{}                 \\
\multicolumn{1}{c|}{\begin{tabular}[c]{@{}c@{}}H$_3$S - $Im\bar{3}m$\\ (250 GPa)\end{tabular}}      & 1.96                 & 879  & 226       & 1.71                 & 817  & 190       \\ \hline \hline
\multicolumn{1}{l|}{}     & \multicolumn{1}{l}{} & \multicolumn{1}{l}{}    & \multicolumn{1}{l|}{}                & \multicolumn{1}{l}{} & \multicolumn{1}{l}{}    & \multicolumn{1}{l}{}                 \\
\multicolumn{1}{c|}{\begin{tabular}[c]{@{}c@{}}ScH$_6$ - $P6_3/mmc$\\  (140 GPa)\end{tabular}}      & 0.99                & 833  & 87.5      & 1.16                 & 728  & 99.6      \\
\multicolumn{1}{l|}{}     & \multicolumn{1}{l}{} & \multicolumn{1}{l}{}    & \multicolumn{1}{l|}{}                & \multicolumn{1}{l}{} & \multicolumn{1}{l}{}    & \multicolumn{1}{l}{}                 \\
\multicolumn{1}{c|}{\begin{tabular}[c]{@{}c@{}}ScH$_6$ - $Cmcm$\\ (100 GPa)\end{tabular}}           & 0.56                & 759  & 20        & 1.36                 & 639  & 108       \\
\multicolumn{1}{l|}{}     & \multicolumn{1}{l}{} & \multicolumn{1}{l}{}    & \multicolumn{1}{l|}{}                & \multicolumn{1}{l}{} & \multicolumn{1}{l}{}    & \multicolumn{1}{l}{}                 \\
\multicolumn{1}{c|}{\begin{tabular}[c]{@{}c@{}}H$_3$S - $R3m$\\  (130 GPa)\end{tabular}}            & 1.79                 & 793  & 175       & 2.10                  & 846  & 214       \\
\multicolumn{1}{l|}{}     & \multicolumn{1}{l}{} & \multicolumn{1}{l}{}    & \multicolumn{1}{l|}{}                & \multicolumn{1}{l}{} & \multicolumn{1}{l}{}    & \multicolumn{1}{l}{}                 \\
\multicolumn{1}{c|}{\begin{tabular}[c]{@{}c@{}}LaBH$_8$ - $Fm\bar{3}m$\\  (100 GPa)\end{tabular}}   & 0.98                 & 819  & 97        & 1.46                 & 729  & 143       \\ 
\multicolumn{1}{l|}{}     & \multicolumn{1}{l}{} & \multicolumn{1}{l}{}    & \multicolumn{1}{l|}{}                & \multicolumn{1}{l}{} & \multicolumn{1}{l}{}    & \multicolumn{1}{l}{}  \\
\multicolumn{1}{c|}{\begin{tabular}[c]{@{}c@{}}H - $Cmca$-4\\  (450 GPa)\end{tabular}}         & 0.89                 & 1742  & 109       & 2.00                 & 1650  & 258       \\
\multicolumn{1}{l|}{}     & \multicolumn{1}{l}{} & \multicolumn{1}{l}{}    & \multicolumn{1}{l|}{}                & \multicolumn{1}{l}{} & \multicolumn{1}{l}{}    & \multicolumn{1}{l}{}                 \\\hline\hline

\end{tabular}
\end{table}

\section{Bonding Analysis}

\begin{table}[]
\caption{Distances and integrated crystal orbital bond indices (iCOBIs) for the listed atom pairs in the Type I compounds for which iCOBI~$\ge$~0.05. The results were the same for the classical and quantum lattice for all of the compounds, except for LaH$_{10}$ where the two sets of values, respectively, are separated by a forward slash.}
\begin{tabular}{l|c|cc}
\hline\hline
Structure                                  & Atom Pair      & \begin{tabular}[c]{@{}c@{}}Distance (\AA)\end{tabular} & iCOBI     \\ \hline
\multirow{2}{*}{PdH - $Fm\bar{3}m$}        & Pd-Pd     & 2.922 & 0.135         \\
                                           & Pd-H      & 2.066 & 0.138         \\ \hline
\multirow{1}{*}{AlH$_3$ - $Pm\bar{3}n$}    & Al-H      & 1.678 & 0.161         \\ \hline
\multirow{4}{*}{LaH$_{10}$ - $Fm\bar{3}m$} & La - H    & 2.096 / 2.096   & 0.122 / 0.091 \\
                                           & La - H    & 2.053 / 2.015 & 0.115 / 0.145 \\
                                           & H - H     & 0.972 / 1.087   & 0.120 / 0.106 \\
                                           & H - H     & 1.254 / 1.165   & 0.055 / 0.072 \\ \hline
\multirow{1}{*}{H - $I4_1/amd$}            & H - H     & 0.985 & 0.144       \\ \hline
PtH - $P6_3/mmc$                           & Pt - H    & 1.938 & 0.064       \\ \hline
\multirow{2}{*}{YH$_6$ - $Im\bar{3}m$}     & Y - H     & 2.017 & 0.124       \\
                                           & H - H     & 1.275 & 0.053       \\  \hline
\multirow{1}{*}{H$_3$S - $Im\bar{3}m$}     & H - S     & 1.534 & 0.347       \\\hline \hline

\end{tabular}
\end{table}

\begin{table}[]
\caption{The distances and integrated crystal orbital bond indices (iCOBIs) for the listed atom pairs in the Type II compounds calculated for both the classic and quantum lattices for which iCOBI~$\ge$~0.05.}
\begin{tabular}{l|c|cc|cc}
\hline \hline
\multicolumn{1}{c}{\multirow{2}{*}{Structure}} & \multicolumn{1}{|c}{\multirow{2}{*}{Atom Pair}} & \multicolumn{2}{|c|}{Classic} & \multicolumn{2}{|c}{Quantum}\\
\multicolumn{1}{c}{}                           & \multicolumn{1}{|c|}{}                           &   Distance (\AA)  & ICOBI   &   Distance (\AA)  & ICOBI  \\ \hline
\multirow{4}{*}{ScH$_6$ - $P6_3/mmc$}    & H-H    & 1.020   & 0.240 & 1.068 & 0.220     \\
                                         & Sc-H   & 1.777   & 0.178 & 1.783 & 0.178     \\
                                         & Sc-H   & 1.896   & 0.137 & 1.884 & 0.141     \\ 
                                         & Sc-Sc  & 2.897   & 0.091 & 2.910 & 0.086     \\ \hline
\multirow{7}{*}{ScH$_6$ - $Cmcm$}        & H-H    & 0.950   & 0.330 & 1.083 & 0.233     \\
                                         & H-H    & 1.275   & 0.117 & 1.083 & 0.233     \\
                                         & Sc-H   & 1.826(5) & 0.169 & 1.833 & 0.174    \\
                                         & Sc-H   & 1.826(8) & 0.195 & 1.833 & 0.174    \\
                                         & Sc-H   & 1.847   & 0.203 & 1.942 & 0.138     \\ 
                                         & Sc-H   & 2.032   & 0.102 & 1.942 & 0.138     \\
                                         & Sc-Sc  & 2.935   & 0.092 & 2.986 & 0.084     \\ \hline
\multirow{2}{*}{H$_3$S - $R3m$ }         & H-S    & 1.456 & 0.463 & 1.559  & 0.347       \\
                                         & H-S    & 1.661 & 0.231 & 1.559  & 0.347       \\ \hline
\multirow{3}{*}{LaBH$_8$ - $Fm\bar{3}m$} & B-H      & 1.409 & 0.281 & 1.454 & 0.291       \\
                                         & B-La     & 2.788  & 0.164 & 2.788  & 0.150       \\
                                         & La-H     & 2.285 & 0.094 & 2.281      & 0.097       \\  \hline
\multirow{3}{*}{H - $Cmca$-4}            & H-H   & 0.782 & 0.424 & 0.832      & 0.358       \\
                                         & H-H   & 1.034 & 0.109 & 1.071      & 0.091       \\
                                         & H-H   & 1.158 & 0.051 & 1.079      & 0.086       \\ \hline \hline
\end{tabular}
\end{table}

\begin{table}[]
\caption{The symmetry inequivalent atoms and their Wyckoff positions in the Type I and Type II structures along with the corresponding symmetry index computed with the iCOHP instead of the iCOBI, $\mathbf{V}(\text{iCOHP})_{x} = \frac{1}{B_x}\sum_{\alpha=1}^{B_x}\text{iCOHP}(x,\alpha)\mathbf{i}_{x\alpha}$ and $S(\text{iCOHP})_x=|\mathbf{V}(\text{iCOHP})_{x}|$(see main text), treating the nuclei as classical, ${S(\text{iCOHP})}_a^\text{c}$,  and quantum, ${S(\text{iCOHP})}_a^\text{q}$, objects. The iCOHP based results are not able to distinguish between Type I and Type II structures as the value of ${S(\text{iCOHP})}_a^\text{c}$ for the hydrogens at the 32f site in LaH$_{10}$ is greater than the Type II H - $Cmca$-4 structure. The Wyckoff parameters given in parenthesis for ScH$_6$ - $Cmcm$ correspond to the symmetry site and multiplicity of the equivalent atoms in the higher symmetry ScH$_6$ - $P6_3/mmc$ phase. Additionally, the values of the iCOHP used in the analysis have been selected using a cutoff of 0.5 (eV/bond), so as to consider the same amount of interactions reported in the iCOBI analysis. }
\begin{tabular}{|ccccc|}
\hline
\multicolumn{1}{|c|}{Structure}                                  & \multicolumn{1}{c|}{Atom} & \multicolumn{1}{c|}{\begin{tabular}[c]{@{}c@{}}Wyckoff\\ letter\end{tabular}} & \multicolumn{1}{c|}{$S(\text{iCOHP})_a^\text{c}$ (eV/bond)} & $S(\text{iCOHP})_a^\text{q}$ (eV/bond)\\ \hline
\multicolumn{5}{|c|}{\multirow{2}{*}{Type I}}                \\
\multicolumn{5}{|c|}{}                                       \\ \hline
\multicolumn{1}{|c|}{\multirow{2}{*}{PdH - $Fm\bar{3}m$}}        & \multicolumn{1}{c|}{Pd}   & \multicolumn{1}{c|}{4a}    & \multicolumn{1}{c|}{0}       & 0       \\ \cline{2-5} 
\multicolumn{1}{|c|}{}                                           & \multicolumn{1}{c|}{H}    & \multicolumn{1}{c|}{4b}    & \multicolumn{1}{c|}{0}       & 0       \\ \hline
\multicolumn{1}{|c|}{\multirow{2}{*}{AlH$_3$ - $Pm\bar{3}n$}}    & \multicolumn{1}{c|}{Al}   & \multicolumn{1}{c|}{2a}    & \multicolumn{1}{c|}{0}       & 0       \\ \cline{2-5} 
\multicolumn{1}{|c|}{}                                           & \multicolumn{1}{c|}{H}    & \multicolumn{1}{c|}{6c}    & \multicolumn{1}{c|}{0}       & 0       \\ \hline
\multicolumn{1}{|c|}{\multirow{3}{*}{LaH$_{10}$ - $Fm\bar{3}m$}} & \multicolumn{1}{c|}{La}   & \multicolumn{1}{c|}{4b}    & \multicolumn{1}{c|}{0.00}    & 0       \\ \cline{2-5} 
\multicolumn{1}{|c|}{}                                           & \multicolumn{1}{c|}{H}    & \multicolumn{1}{c|}{8c}    & \multicolumn{1}{c|}{0.01}    & 0.01    \\ \cline{2-5} 
\multicolumn{1}{|c|}{}                                           & \multicolumn{1}{c|}{H}    & \multicolumn{1}{c|}{32f}   & \multicolumn{1}{c|}{0.22}    & 0.08  \\ \hline
\multicolumn{1}{|c|}{H - $I4_1/amd$}                             & \multicolumn{1}{c|}{H}    & \multicolumn{1}{c|}{4b}    & \multicolumn{1}{c|}{0}       & 0       \\ \hline
\multicolumn{1}{|c|}{\multirow{2}{*}{PtH - $P6_3/mmc$}}       & \multicolumn{1}{c|}{Pt}   & \multicolumn{1}{c|}{2d}    & \multicolumn{1}{c|}{0}          & 0       \\ \cline{2-5} 
\multicolumn{1}{|c|}{}                                           & \multicolumn{1}{c|}{H}    & \multicolumn{1}{c|}{2a}    & \multicolumn{1}{c|}{0}       & 0       \\ \hline
\multicolumn{1}{|c|}{\multirow{2}{*}{YH$_6$ - $Im\bar{3}m$}}     & \multicolumn{1}{c|}{Y}    & \multicolumn{1}{c|}{2a}    & \multicolumn{1}{c|}{0}       & 0       \\ \cline{2-5} 
\multicolumn{1}{|c|}{}                                           & \multicolumn{1}{c|}{H}    & \multicolumn{1}{c|}{12d}   & \multicolumn{1}{c|}{0}       & 0       \\ \hline
\multicolumn{1}{|c|}{\multirow{2}{*}{H$_3$S - $Im\bar{3}m$}}     & \multicolumn{1}{c|}{S}    & \multicolumn{1}{c|}{2a}    & \multicolumn{1}{c|}{0}       & 0       \\ \cline{2-5} 
\multicolumn{1}{|c|}{}                                           & \multicolumn{1}{c|}{H}    & \multicolumn{1}{c|}{6b}    & \multicolumn{1}{c|}{0}       & 0       \\ \hline
\multicolumn{5}{|c|}{\multirow{2}{*}{Type II}}                \\
\multicolumn{5}{|c|}{}                                       \\ \hline
\multicolumn{1}{|c|}{\multirow{2}{*}{ScH$_6$ - $P6_3/mmc$}}   & \multicolumn{1}{c|}{Sc}   & \multicolumn{1}{c|}{2d}    & \multicolumn{1}{c|}{0}        &    0     \\ \cline{2-5} 
\multicolumn{1}{|c|}{}                                           & \multicolumn{1}{c|}{H}    & \multicolumn{1}{c|}{12k}   & \multicolumn{1}{c|}{0.52}  & 0.48  \\ \hline
\multicolumn{1}{|c|}{\multirow{3}{*}{ScH$_6$ - $Cmcm$}}          & \multicolumn{1}{c|}{Sc}   & \multicolumn{1}{c|}{4c}    & \multicolumn{1}{c|}{0.00}  & 0 (2d)      \\ \cline{2-5} 
\multicolumn{1}{|c|}{}                                           & \multicolumn{1}{c|}{H}    & \multicolumn{1}{c|}{16h}   & \multicolumn{1}{c|}{0.33}  & 0.47 (12k) \\ \cline{2-5} 
\multicolumn{1}{|c|}{}                                           & \multicolumn{1}{c|}{H}    & \multicolumn{1}{c|}{8f}    & \multicolumn{1}{c|}{0.77}  & 0.47 (12k) \\ \hline
\multicolumn{1}{|c|}{\multirow{2}{*}{H$_3$S - $R3m$}}            & \multicolumn{1}{c|}{S}    & \multicolumn{1}{c|}{3a}    & \multicolumn{1}{c|}{0.68}  & 0       \\ \cline{2-5} 
\multicolumn{1}{|c|}{}                                           & \multicolumn{1}{c|}{H}    & \multicolumn{1}{c|}{9b}    & \multicolumn{1}{c|}{1.03}  & 0       \\ \hline
\multicolumn{1}{|c|}{\multirow{3}{*}{LaBH$_8$ - $Fm\bar{3}m$}}   & \multicolumn{1}{c|}{La}   & \multicolumn{1}{c|}{4b}    & \multicolumn{1}{c|}{0}       & 0       \\ \cline{2-5} 
\multicolumn{1}{|c|}{}                                           & \multicolumn{1}{c|}{B}    & \multicolumn{1}{c|}{4a}    & \multicolumn{1}{c|}{0}       & 0       \\ \cline{2-5} 
\multicolumn{1}{|c|}{}                                           & \multicolumn{1}{c|}{H}    & \multicolumn{1}{c|}{32f}   & \multicolumn{1}{c|}{0.70}  & 0.73  \\ \hline
\multicolumn{1}{|c|}{H - $Cmca$-4}                               & \multicolumn{1}{c|}{H}    & \multicolumn{1}{c|}{8f}    & \multicolumn{1}{c|}{0.17}  & 0.09  \\ \hline
\end{tabular}
\end{table}

\begin{table}[]
\caption{The symmetry inequivalent atoms and their Wyckoff positions in the Type I and Type II structures along with the corresponding symmetry index computed with the magnitudes of the interatomic distances, $\mathbf{V}(\text{d})_{x} = \frac{1}{B_x}\sum_{\alpha=1}^{B_x}\text{d}(x,\alpha)\mathbf{i}_{x\alpha}$ and $S(\text{d})_x=|\mathbf{V}(\text{d})_{x}|$(see main text), treating the nuclei as classical, ${S(\text{d})}_a^\text{c}$,  and quantum, ${S(\text{d})}_a^\text{q}$, objects. Here $\text{d}(x,\alpha)$ is the interatomic distance between atoms $x$ and $\alpha$. The distance based results are not able to distinguish between Type I and Type II structures as the value of ${S(\text{d})}_a^\text{c}$ for the hydrogens at the 32f site in LaH$_{10}$ is greater than the ${S(\text{d})}_a^\text{c}$ for both atoms in the Type II H$_3$S - $R3m$ structure. The Wyckoff parameters given in parenthesis for ScH$_6$ - $Cmcm$ correspond to the symmetry site and multiplicity of the equivalent atoms in the higher symmetry ScH$_6$ - $P6_3/mmc$ phase. The interatomic distances considered in this analysis correspond to those considered for the iCOBI analysis.
}
\begin{tabular}{|ccccc|}
\hline
\multicolumn{1}{|c|}{Structure}                                  & \multicolumn{1}{c|}{atom} & \multicolumn{1}{c|}{\begin{tabular}[c]{@{}c@{}}Wyckoff\\ letter\end{tabular}} & \multicolumn{1}{c|}{$S(\text{d})_a^c$ (\AA\ )}  & $S(\text{d})_a^q$ (\AA\ )\\ \hline
\multicolumn{5}{|c|}{\multirow{2}{*}{Type 1}}                \\
\multicolumn{5}{|c|}{}                                       \\ \hline
\multicolumn{1}{|c|}{\multirow{2}{*}{PdH - $Fm\bar{3}m$}}        & \multicolumn{1}{c|}{Pd}   & \multicolumn{1}{c|}{4a}    & \multicolumn{1}{c|}{0}       & 0       \\ \cline{2-5} 
\multicolumn{1}{|c|}{}                                           & \multicolumn{1}{c|}{H}    & \multicolumn{1}{c|}{4b}    & \multicolumn{1}{c|}{0}       & 0       \\ \hline
\multicolumn{1}{|c|}{\multirow{2}{*}{AlH$_3$ - $Pm\bar{3}n$}}    & \multicolumn{1}{c|}{Al}   & \multicolumn{1}{c|}{2a}    & \multicolumn{1}{c|}{0}       & 0       \\ \cline{2-5} 
\multicolumn{1}{|c|}{}                                           & \multicolumn{1}{c|}{H}    & \multicolumn{1}{c|}{6c}    & \multicolumn{1}{c|}{0}       & 0       \\ \hline
\multicolumn{1}{|c|}{\multirow{3}{*}{LaH$_{10}$ - $Fm\bar{3}m$}} & \multicolumn{1}{c|}{La}   & \multicolumn{1}{c|}{4b}    & \multicolumn{1}{c|}{0}       & 0       \\ \cline{2-5} 
\multicolumn{1}{|c|}{}                                           & \multicolumn{1}{c|}{H}    & \multicolumn{1}{c|}{8c}    & \multicolumn{1}{c|}{0.00}    & 0       \\ \cline{2-5} 
\multicolumn{1}{|c|}{}                                           & \multicolumn{1}{c|}{H}    & \multicolumn{1}{c|}{32f}   & \multicolumn{1}{c|}{0.18}    & 0.03  \\ \hline
\multicolumn{1}{|c|}{H - $I4_1/amd$}                             & \multicolumn{1}{c|}{H}    & \multicolumn{1}{c|}{4b}    & \multicolumn{1}{c|}{0}       & 0       \\ \hline
\multicolumn{1}{|c|}{\multirow{2}{*}{PtH - $P6_3/mmc$}}          & \multicolumn{1}{c|}{Pt}   & \multicolumn{1}{c|}{2d}    & \multicolumn{1}{c|}{0}       & 0       \\ \cline{2-5} 
\multicolumn{1}{|c|}{}                                           & \multicolumn{1}{c|}{H}    & \multicolumn{1}{c|}{2a}    & \multicolumn{1}{c|}{0}       & 0       \\ \hline
\multicolumn{1}{|c|}{\multirow{2}{*}{YH$_6$ - $Im\bar{3}m$}}     & \multicolumn{1}{c|}{Y}    & \multicolumn{1}{c|}{2a}    & \multicolumn{1}{c|}{0}       & 0       \\ \cline{2-5} 
\multicolumn{1}{|c|}{}                                           & \multicolumn{1}{c|}{H}    & \multicolumn{1}{c|}{12d}   & \multicolumn{1}{c|}{0}       & 0       \\ \hline
\multicolumn{1}{|c|}{\multirow{2}{*}{H$_3$S - $Im\bar{3}m$}}     & \multicolumn{1}{c|}{S}    & \multicolumn{1}{c|}{2a}    & \multicolumn{1}{c|}{0}       & 0       \\ \cline{2-5} 
\multicolumn{1}{|c|}{}                                           & \multicolumn{1}{c|}{H}    & \multicolumn{1}{c|}{6b}    & \multicolumn{1}{c|}{0}       & 0       \\ \hline
\multicolumn{5}{|c|}{\multirow{2}{*}{Type 2}}                \\
\multicolumn{5}{|c|}{}                                       \\ \hline
\multicolumn{1}{|c|}{\multirow{2}{*}{ScH$_6$ - $P6_3/mmc$}}      & \multicolumn{1}{c|}{Sc}   & \multicolumn{1}{c|}{2d}    & \multicolumn{1}{c|}{0}       & 0       \\ \cline{2-5} 
\multicolumn{1}{|c|}{}                                           & \multicolumn{1}{c|}{H}    & \multicolumn{1}{c|}{12k}   & \multicolumn{1}{c|}{0.29}    & 0.32  \\ \hline
\multicolumn{1}{|c|}{\multirow{3}{*}{ScH$_6$ - $Cmcm$}}          & \multicolumn{1}{c|}{Sc}   & \multicolumn{1}{c|}{4c}    & \multicolumn{1}{c|}{0.38}    & 0  (2d)     \\ \cline{2-5} 
\multicolumn{1}{|c|}{}                                           & \multicolumn{1}{c|}{H}    & \multicolumn{1}{c|}{16h}   & \multicolumn{1}{c|}{0.41}    & 0.32 (12k) \\ \cline{2-5} 
\multicolumn{1}{|c|}{}                                           & \multicolumn{1}{c|}{H}    & \multicolumn{1}{c|}{8f}    & \multicolumn{1}{c|}{0.26}    & 0.32 (12k) \\ \hline
\multicolumn{1}{|c|}{\multirow{2}{*}{H$_3$S - $R3m$}}            & \multicolumn{1}{c|}{S}    & \multicolumn{1}{c|}{3a}    & \multicolumn{1}{c|}{0.02}    & 0       \\ \cline{2-5} 
\multicolumn{1}{|c|}{}                                           & \multicolumn{1}{c|}{H}    & \multicolumn{1}{c|}{9b}    & \multicolumn{1}{c|}{0.11}    & 0       \\ \hline
\multicolumn{1}{|c|}{\multirow{3}{*}{LaBH$_8$ - $Fm\bar{3}m$}}   & \multicolumn{1}{c|}{La}   & \multicolumn{1}{c|}{4b}    & \multicolumn{1}{c|}{0}       & 0       \\ \cline{2-5} 
\multicolumn{1}{|c|}{}                                           & \multicolumn{1}{c|}{B}    & \multicolumn{1}{c|}{4a}    & \multicolumn{1}{c|}{0}       & 0       \\ \cline{2-5} 
\multicolumn{1}{|c|}{}                                           & \multicolumn{1}{c|}{H}    & \multicolumn{1}{c|}{32f}   & \multicolumn{1}{c|}{0.20}    & 0.25  \\ \hline
\multicolumn{1}{|c|}{H - $Cmca$-4}                               & \multicolumn{1}{c|}{H}    & \multicolumn{1}{c|}{8f}    & \multicolumn{1}{c|}{0.29}    & 0.22  \\ \hline
\end{tabular}
\end{table}

\end{document}